\documentclass[12pt]{article}

\usepackage{newtxtext,newtxmath}

\usepackage{graphicx}

\usepackage[letterpaper,margin=1in]{geometry}

\renewenvironment{abstract}
	{\quotation}
	{\endquotation}

\date{}

\makeatletter
\renewcommand{\fnum@figure}{\textbf{Figure \thefigure}}
\renewcommand{\fnum@table}{\textbf{Table \thetable}}
\makeatother

\usepackage[numbers,sort&compress]{natbib}
\setcitestyle{numbers,round,comma}

\makeatletter
\renewcommand{\@biblabel}[1]{#1.}
\makeatother

\usepackage{url}

\usepackage{lineno}
\usepackage[hidelinks]{hyperref}

\def\scititle{Tracing the space-time causal origins of Earth system extremes}
\title{\bfseries \boldmath \scititle}

\author{
	Jhayron S. Pérez-Carrasquilla$^{1\ast}$,
	J. Jake Nichol$^{2}$,
	Vanessa Robledo$^{3},$\and
    Diana Bull$^{2}$, 
    Katherine Dagon$^{4}$, 
    Michael N. Evans$^{5,6}$,
    and Maria J. Molina$^{1}$\and
	\small$^{1}$Department of Atmospheric and Oceanic Science, University of Maryland, College Park, MD, USA\and
	\small$^{2}$Sandia National Laboratories, Albuquerque, NM, USA\and
	\small$^{3}$IIHR—Hydroscience \& Engineering, University of Iowa, Iowa City, IA, USA\and
	\small$^{4}$NSF National Center for Atmospheric Research, Boulder, CO, USA\and
    \small$^{5}$Department of Geological, Environmental and Planetary Sciences, University of Maryland,\and 
    \small{College Park, MD, USA}\and
    \small$^{6}$Earth System Science Interdisciplinary Center, University of Maryland, College Park, MD, USA \and
	\small$^\ast$Corresponding author. Email: jhayron@umd.edu\and
}

\begin{document} 
\maketitle


\begin{abstract} \bfseries \boldmath
Identifying the causes of Earth's extremes is challenging because counterfactual experiments are not possible in the observed world. Data-driven causal discovery complements computationally expensive and potentially biased numerical experiments, but existing methods can struggle with undersampled, high-dimensional data and fail to recover multi-timestep, multivariate pathways leading to specific events. We introduce Tracer of Causal Evolutions in Space and Time (TraCE-ST), a probabilistic Lagrangian approach that produces event-conditioned causal trajectories in multivariate gridded data. TraCE-ST recovers known causal drivers and estimates their relative contributions in synthetic experiments and real-world extremes, including the 1991 Mount Pinatubo eruption. TraCE-ST also highlights less-studied drivers, such as orography-driven vorticity for Tropical Storm Debby (2006) and anomalous ocean-surface fluxes for the 2021 Pacific Northwest heatwave. Here, we propose causal tracking as an efficient data-driven framework for synthesizing causal evidence and generating testable hypotheses, complementing association analyses and numerical modeling while accelerating the study of high-impact events.
\end{abstract}

\newpage

\noindent\textbf{Short title:} Tracing causal origins of Earth system extremes

\noindent\textbf{Teaser:} A causal tracking framework uncovers the space-time pathways leading to high-impact extreme events.

\section*{Introduction}

Discovering causal dependencies in high-dimensional dynamical systems provides the mechanistic foundation for explaining and predicting complex physical phenomena. This task is especially challenging in Earth systems, where weather and climate extremes emerge from multiscale interactions, nonlinear feedbacks, and nonstationary dynamics \cite{runge_detecting_2019,nathaniel_deep_2025}, with broad socioeconomic, biological, and ecological implications. Explaining why a particular event occurred requires more than detecting statistical association \cite{runge_inferring_2019}; it requires identifying the variables and processes that shaped the sequence of states leading to a specific outcome \cite{friedman_explanation_1974, Pearl2009}. To achieve this, we often rely on numerical models, but they are computationally expensive and prone to biases \cite{runge_detecting_2019}. Causal discovery methods, such as Granger causality, Peter–Clark Momentary Conditional Independence (PCMCI) and its extensions, DYNOTEARS, and deep Koopman-based approaches, offer an alternative to address this challenge by inferring directed causal dependencies from multivariate time series \cite{runge_detecting_2019,mcgraw_memory_2018,ebert-uphoff_causal_2012,pamfil_dynotears_2020,runge_causal_2023,tibau_spatiotemporal_2022,gao2024causaldiscoverysemistationarytime,rupe2025causaldiscoverynonlineardynamical,nathaniel_deep_2025}. However, these methods are generally designed to recover causal structure under assumptions about observability, sampling, temporal ordering, and, in many cases, stationarity, that make them challenging to use in high-dimensional settings with few temporal samples. Furthermore, they do not produce event-specific explanations of the realized causal chain leading to an outcome \cite{runge_causal_2023}.

Recent methods, including Causal Space-Time Stencil Learning (CaStLe) and its multivariate version (M-CaStLe), extend causal discovery to high-dimensional gridded systems by exploiting spatial regularities to estimate causal dependencies in relatively short temporal records \cite{nichol_space-time_2025,nichol_mcastle_2026}. These methods reduce the sample-sparsity barriers that arise in Earth-system data, where the number of spatial locations and variables can far exceed the number of temporal observations. However, local causal discovery and event explanation are distinct tasks. M-CaStLe identifies directed dependencies among variables and nearby states ($t-1 \rightarrow t$), but it does not directly explain how a particular event emerged through a realized sequence of interacting processes. Earth system events often arise from extended, multi-step pathways in which local and remote drivers interact across space, time, and variables. The remaining challenge is therefore to move from local causal structure to event-conditioned causal pathways that connect upstream processes to a realized event while quantifying the relative contributions of co-occurring and competing drivers. This event-conditioned pathway framing is increasingly important for Earth system science as many weather and climate extremes compound, intensify, or become more frequent under continued warming \cite{zscheischler_future_2018,seneviratne2021weather}.

To address this challenge, we introduce the Tracer of Causal Evolutions in Space and Time (TraCE-ST), a Lagrangian-inspired framework for identifying event-conditioned causal pathways among interacting Earth system variables. TraCE-ST is motivated by Lagrangian trajectory methods used across Earth system science, which trace the histories of air parcels, water parcels, particles, or other transported quantities backward or forward in time using spatiotemporal fields to identify upstream sources, transport pathways, and process histories \cite{stohl_technical_2005, stein_noaas_2015, sebille_lagrangian_2018, molina_moisture_2019, perez-carrasquilla_forecasting_2023,velasquez-garcia_long-range_2024}. Analogously, TraCE-ST traces inferred causal influence backward through localized directed graphs rather than tracing physical parcels through a velocity field: TraCE-ST recursively follows inferred causal parents across successive time steps and outputs the multi-step chain of interactions that precede the realized state of a target variable within the studied system. This framework can also flexibly accommodate different causal discovery engines, provided that they return directed temporal dependencies suitable for pathway tracing. A probabilistic TraCE-ST formulation further enables estimation of co-occurring causal pathways and their relative contributions. More generally, TraCE-ST provides a systematic way to trace cross-variable causal pathways among interacting physical fields, identify remote influences mediated by local interactions, and estimate the relative importance of distinct processes influencing a local outcome. 

We validate TraCE-ST hierarchically across systems of increasing complexity, including two synthetic cases and three real-world scenarios: (i) the development of Tropical Storm Debby in 2006 \cite{lin_origin_2013,chen_relation_2014}, (ii) volcanic aerosol dispersion following the 1991 Mount Pinatubo eruption \cite{mccormick_atmospheric_1995}, and (iii) the formation of the 2021 North American Pacific Northwest (PNW) heatwave \cite{white_unprecedented_2023}. Across these applications, TraCE-ST outputs causal pathways consistent with established mechanisms while also shedding light on less-studied candidate drivers. By transforming causal discovery from local graph estimation into dynamic, event-centered pathway tracing, TraCE-ST offers a new systematic framework for studying the causal evolution of extreme events in Earth systems.

\section*{Results}
\subsection*{TraCE-ST: Tracing Causal Evolutions in Space and Time}
 
\begin{figure}[t!]
	\centering
	\includegraphics[width=0.56\textwidth]{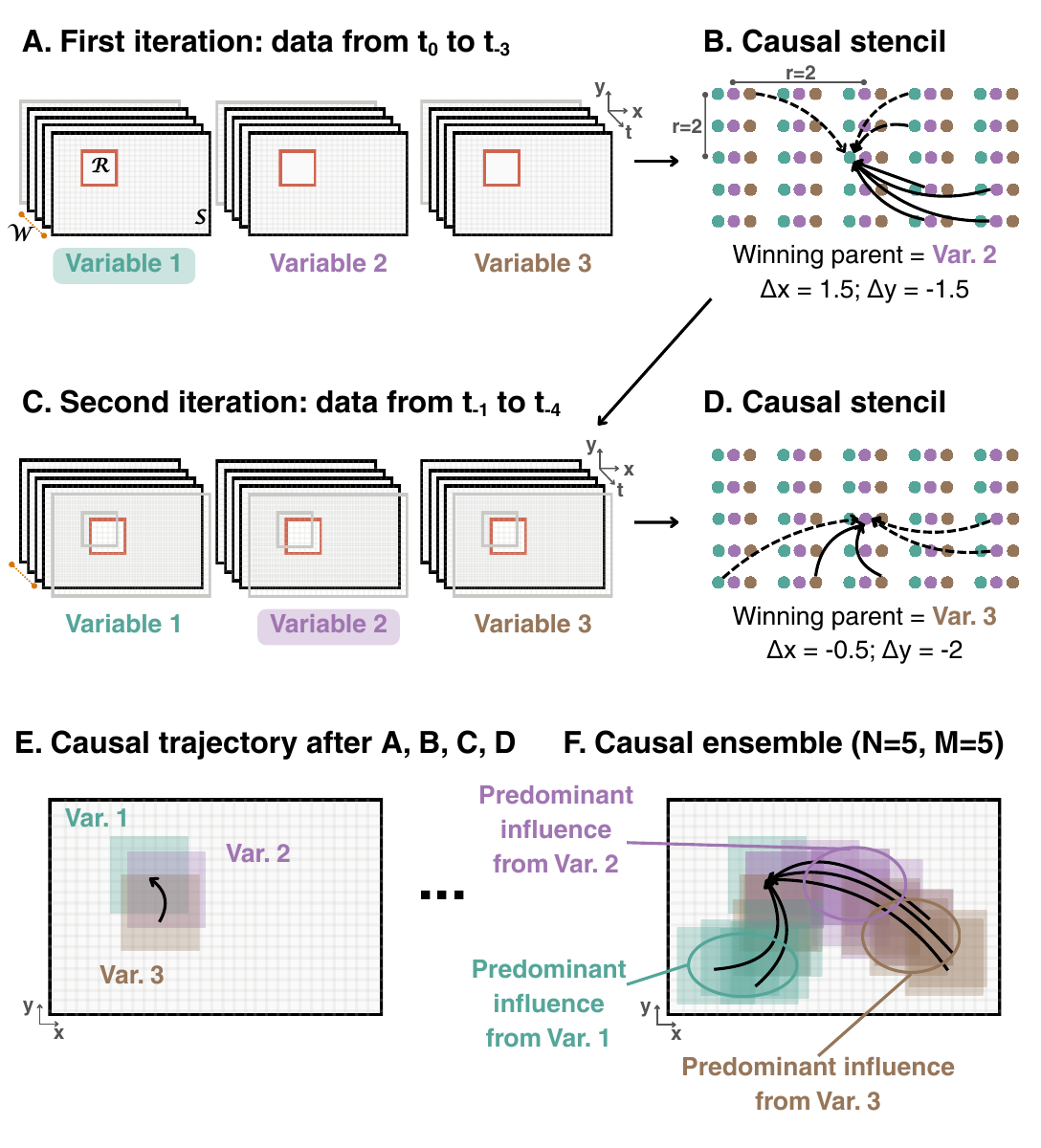}
    \caption{\textbf{Illustration of TraCE-ST using an idealized system with three ($p=3$) variables defined on a two-dimensional ($x, y$) spatial grid.}
        (\textbf{A}, \textbf{C}) Input data provided to M-CaStLe at the first and second tracing iterations. Orange polygons denote the analysis region $\mathcal{R}$, which spans $7\times7$ grid cells in this example. Orange dashed lines across the black-outlined domains indicate the temporal window $\mathcal{W}$, with length $L=4$. Highlighted labels indicate the child variable at each step: Variable 1 at $t_0$ and Variable 2 at $t_{-1}$. In (\textbf{C}), both $\mathcal{R}$ and $\mathcal{W}$ have shifted relative to their initial locations in (\textbf{A}); previous locations are indicated with grey polygons.
        (\textbf{B}, \textbf{D}) Causal stencil $\mathcal{B}_r(s)$, with radius $r=2$, and the corresponding set of candidate parents $\mathrm{P}_{\mathcal{W},\mathcal{R}}$ for the current child variable. Colors denote variables. Arrows indicate statistically significant causal parents from $t-1$ to $t$ according to M-CaStLe, and solid arrows identify the ``winning'' parent cluster that determines the next child variable and the updated position of $\mathcal{R}$ at $t-1$.
        (\textbf{E}) Resulting causal trajectory $\tau$ after two iterations, where Variable 2 at $t_{-1}$ and Variable 3 at $t_{-2}$ are identified as winning causal parents.
        (\textbf{F}) Ensemble of causal trajectories $\Psi$ ($N=5$ tracing steps, $M=5$ ensemble members). Locations and variables that are more frequently selected as causal parents of the target child $C^{(j_0)}(\mathcal{R}_0,t_0)$ are indicated.}
	\label{fig:trace_st_schematic}
\end{figure}

As discussed above, TraCE-ST builds upon localized causal graphs estimated with M-CaStLe (see Materials and Methods) \cite{nichol_mcastle_2026} and outputs trajectories that represent the temporal sequence of multivariate contributions leading to a target event. Figure \ref{fig:trace_st_schematic} provides a schematic overview of TraCE-ST, illustrating the successive steps of local causal discovery, grouping of causal parents, and backward Lagrangian-inspired propagation through the analysis domain.

Let a multivariate gridded anomaly field be denoted by $\mathbf{X}$, with $\mathbf{X}(s,t)= ( X^{(1)}(s,t), \dots, X^{(p)}(s,t))$, where $p$ is the number of variables, $s=(x,y)$ is a grid location in the spatial domain $\mathcal{S}$, and $t$ is a time index in the temporal domain $\mathcal{T}$. For a selected variable $j$, where $j\in\{1,\dots,p\}$, the local value of that variable at location $s$ and time $t$ is denoted by $X^{(j)}(s,t) \in \mathbb{R}$. For a given analysis region $\mathcal{R}$ within $\mathcal{S}$ at time $t$, we define the `child' $C^{(j)}(\mathcal{R},t)$ (i.e., explained effect) as the spatial mean of anomaly values $X^{(j)}(s,t)$ over all grid cells $s=(x,y)$ contained in $\mathcal{R}$:
\begin{equation}
C^{(j)}(\mathcal{R},t) :=
\frac{1}{|\mathcal{R}|}
\sum_{s \in \mathcal{R}}
X^{(j)}(s,t)
\label{eq:event_definition}
\end{equation}
where $|\mathcal{R}|$ denotes the number of spatial grid elements in $\mathcal{R}$. A spatial mean is used in this definition to remain consistent with M-CaStLe, which estimates causal relationships over an analysis region rather than for individual grid cells. The `target child,' $C^{(j_0)}(\mathcal{R}_0,t_0)$, represents the weather or climate event to be studied—where $j_0$, $\mathcal{R}_0$, and $t_0$ denote the selected variable, analysis region, and time of interest—and serves as the starting point for TraCE-ST trajectory construction.

For a given child $C^{(j)}(\mathcal{R},t)$, causal parents are estimated over a temporal window $\mathcal{W}=\{t-L+1,\dots,t\}$, where $L$ is the number of temporal samples in the window. M-CaStLe \cite{nichol_mcastle_2026} assumes that causal relationships are approximately stationary over the respective temporal window $\mathcal{W}$ and spatially homogeneous over the analysis region $\mathcal{R}$. These assumptions are most appropriate when $\mathcal{W}$ and $\mathcal{R}$ are chosen such that the underlying interactions vary only weakly in time and space. Candidate parents are restricted to a single temporal lag, from $t-1$ to $t$, and to local spatial stencils centered on grid locations within $\mathcal{R}$. Specifically, for each $s\in\mathcal{R}$, let $\mathcal{B}_r(s)$ denote the radius-$r$ spatial stencil centered on $s$. Candidate parents (i.e., potential causes) take the form $X^{(a)}(s',t-1)$, where $s' \in \mathcal{B}_r(s)$ for some $s \in \mathcal{R}$, and $a \in \{1,\dots,p\}$, which also includes the variable $j$. Thus, the inferred set of causal parents for child $C^{(j)}(\mathcal{R},t)$ within the analysis region $\mathcal{R}$ and temporal window $\mathcal{W}$ satisfies
\begin{equation}
\mathrm{P}_{\mathcal{W},\mathcal{R}}
\subseteq
\Big\{
X^{(a)}(s',t-1)
:
s \in \mathcal{R},
s' \in \mathcal{B}_r(s);
a \in \{1,\dots,p\}
\Big\}
\label{eq:parents_set_definition}
\end{equation}
At each grid location $s$, this statement implies at most $p\times|\mathcal{B}_r(s)| = p\times(2r+1)^2$ admissible causal parents for $C^{(j)}(\mathcal{R},t)$.

Within M-CaStLe, an underlying time-series graph discovery algorithm (e.g., Granger causality, PCMCI, or DYNOTEARS) identifies statistically significant causal parents based on predictive power or conditional independence tests. Once the set of causal parents $\mathrm{P}_{\mathcal{W},\mathcal{R}}$ is identified, these parents are grouped according to the variable, the sign of their inferred causal relationship to the child, and their spatial direction relative to $s$ (the center of the stencil $\mathcal{B}_r(s)$). The DBSCAN algorithm (density-based spatial clustering) \cite{schubert_dbscan_2017} is used to identify these parent groups (see Materials and Methods). This parent-grouping step aggregates parents that represent coherent propagation patterns in the stencil space. Here, ``causal strength'' refers to the magnitude returned by the chosen causal discovery engine for each significant parent-child relationship (e.g., regression coefficient or conditional-dependence measure).

At each tracing step of a trajectory, one of the resulting parent groups from the set $\mathrm{P}_{\mathcal{W},\mathcal{R}}$ is then selected to define the next step of the causal tracing procedure. The selection can be performed deterministically or probabilistically. Deterministic selection chooses the DBSCAN cluster of causal parents with either the largest summed causal strength, favoring groups of multiple parents aligned along a common direction, or the largest average causal strength, favoring clusters containing fewer but stronger parent-child relationships (see Materials and Methods). Probabilistic selection instead samples a cluster using probabilities derived from its summed or mean causal strength, calculated using either linear normalization or a softmax function (see Materials and Methods). The latter formulation enables a Monte Carlo exploration of multiple competing causal pathways. The variable associated with the selected parent group becomes the child variable for the next tracing step. The window is also shifted backward in time, becoming $\mathcal{W}=\{t-L,\dots,t-1\}$. Furthermore, the analysis region $\mathcal{R}$ is translated by a displacement vector computed as the weighted mean offset of the selected parent group with respect to the center of $\mathcal{B}_r(s)$, preserving the size of the analysis region $\mathcal{R}$. This procedure, therefore, propagates $\mathcal{R}$ through space and backward in time following causal influences.

Repeating the above procedure iteratively $N$ times produces a causal trajectory defined as
\begin{equation}
\tau := \big\{ Z_{t_0}, Z_{t_{-1}}, \dots, Z_{t_{-N}} \big\},
\label{eq:trajectory_definition}
\end{equation}
where each trajectory element $Z_{t_{-i}}$ corresponds to an ordered pair $(j_{-i},\mathcal{R}_{-i})$, with $t_{-i} = t_0 - i$. This pair represents the traced variable $j$ and the spatial region $\mathcal{R}$ at each time step. For $i>0$, $j_{-i}$ denotes the selected parent variable at lag $i$, which becomes the child variable for the next backward tracing step. These pairs, therefore, indicate the inferred causal origin of the target child $C^{(j_0)}(\mathcal{R}_0,t_0)$ at each time lag. The number of tracing steps, $N$, is application-dependent and can be extended as long as informative causal parents continue to be identified, allowing TraCE-ST to investigate processes operating over a broad range of timescales. However, uncertainty in the inferred causal pathways accumulates, and therefore generally increases backward in time, as causes become more temporally distant from the event.

An ensemble of causal trajectories $\Psi = (\tau_{1}, \tau_{2},\dots,\tau_{M})$ can be generated by repeating the tracing procedure $M$ times using the probabilistic approach for parent group selection. With a sufficiently large $M$, the spatial density of trajectories can be computed for each traced parent variable at each time lag as the frequency with which the corresponding region $\mathcal{R}$ occupies a given location in the domain. This density indicates where and how often each variable is selected as a causal parent during backward tracing from the target child $C^{(j_0)}(\mathcal{R}_0,t_0)$, which can be interpreted as an empirical estimate of the relative contribution of each variable toward causing the target child at $t=0$ (e.g., Fig.~\ref{fig:trace_st_schematic}F). Given the presence of co-occurring causal parents, we recommend interpreting TraCE-ST results using a sufficiently large ensemble rather than individual trajectories.

Several hyperparameters impact the behavior of TraCE-ST, including stencil radius, window size, region extent, and the underlying causal discovery algorithm (see Materials and Methods). Since different hyperparameter combinations can yield distinct plausible probabilistic causal trajectories for the target child $C^{(j_0)}(\mathcal{R}_0,t_0)$, sensitivity should be explored and used to characterize uncertainty. Given this sensitivity, it is important that each TraCE-ST application defines physically motivated exploration ranges for hyperparameters. In the Materials and Methods, we describe each hyperparameter and provide guidance for selecting appropriate ranges. In general, the temporal resolution and spatial grid spacing of the data, TraCE-ST time step, and stencil radius should be chosen so that the relevant physical processes can be resolved at the scales of interest. Likewise, the temporal window and region size should be selected to provide sufficient samples for M-CaStLe while maintaining the assumptions of approximate temporal and spatial stationarity. 

Hyperparameter settings should also be screened using application-specific criteria. In the following applications, within the hyperparameter exploration ranges, we excluded settings that produced trajectories that were too short to resolve the event's evolution (100\% of prescribed trajectory length for the synthetic cases and 80\% for the real-world cases). In Debby and Mount Pinatubo cases, we also excluded settings producing ensembles with a large fraction of members terminating before a prescribed tracing period (more than 33\% of ensemble size). Earth-system constraints can provide additional criteria for excluding hyperparameter ranges that often produce trajectories with unrealistic behavior, such as transitions between variables and locations that are inconsistent with known physics. Additionally, regions of the hyperparameter space that yield substantially different causal trajectories under small changes in neighboring hyperparameter settings should not be considered. The goal should not be to select a single optimal configuration, but to identify causal pathways that are consistent across a range of settings. Accordingly, we restricted our following analyses to ranges of hyperparameter settings that avoided frequent premature trajectory termination or trajectories confined to the same variable or location.

\subsection*{Controlled cases with prescribed causal trajectories}

First, we evaluate TraCE-ST in two controlled synthetic cases. In the first controlled case, the system consists of two variables, $V_1$ and $V_2$. A Gaussian perturbation with a prescribed trajectory propagates across $V_1$, while $V_2$ is correlated with $V_1$ but does not cause it. Consequently, along the prescribed trajectory, the true causal parent of $V_1$ at each time step $t_{-i}$ is $V_1$ at the previous lag $t_{-(i+1)}$, despite the presence of cross-variable correlation. With this configuration, we test whether TraCE-ST can correctly reconstruct the causal propagation of a coherent feature in space and time while avoiding spurious attribution to a correlated variable. In the second controlled case, the system contains three variables, $V_1$, $V_2$, and $V_3$. Two Gaussian perturbations propagate independently along prescribed trajectories in $V_1$ and $V_3$ and collide, producing a response in $V_2$ in which $V_1$ and $V_3$ are assigned different relative contributions using a linear coefficient. This setup mimics situations in which a target child arises from multiple interacting causal parents, allowing us to evaluate whether the probabilistic formulation of TraCE-ST can recover these parents and quantify their relative contributions using the frequency of each as a causal parent.

Both experiments are conducted on a $200\times200$ Cartesian grid with grid spacing of $\Delta x = \Delta y = 1$. Gaussian perturbations are chosen because they provide spatially coherent signals whose propagation can be prescribed precisely while remaining representative of localized anomalies commonly observed in geophysical fields. The perturbations have an amplitude of $1$ and a spatial scale of approximately $20\times20$ grid cells, and propagate along deterministic trajectories for $30$ time steps prior to the event time $t_0$. Note that the spatial size of the prescribed perturbations is large enough to expand over multiple pixels, in line with the assumption of spatial homogeneity from M-CaStLe (see Materials and Methods). To emulate non-causal variability, all variables include spatially coherent autoregressive Gaussian noise of order 1, $\nu$, with temporal autocorrelation of $0.85$ and amplitude of $0.1$. Spatial coherence was introduced by applying a Gaussian spatial filter to the Gaussian noise at each time step before the autoregressive update using a smoothing scale of 1.5 grid cells in both spatial dimensions. These settings produce persistent, spatially organized background fluctuations that obscure the imposed causality without introducing additional cross-variable causality. 

In both cases, TraCE-ST trajectories are computed for $N=30$ backward tracing steps starting from the location of the target child at time $t_0$. Although idealized, the synthetic experiments were designed to reflect key characteristics of the subsequent Earth-system case studies: spatially coherent features spanning multiple grid cells, evolution across multiple time steps, and causal discovery over relatively short temporal windows. The resulting benchmarks therefore evaluate TraCE-ST under application-relevant conditions while retaining fully prescribed ground-truth causal trajectories.

\subsubsection*{Causal reconstruction in a two-variable system}

\begin{figure}[t!]
	\centering
	\includegraphics[width=1\textwidth]{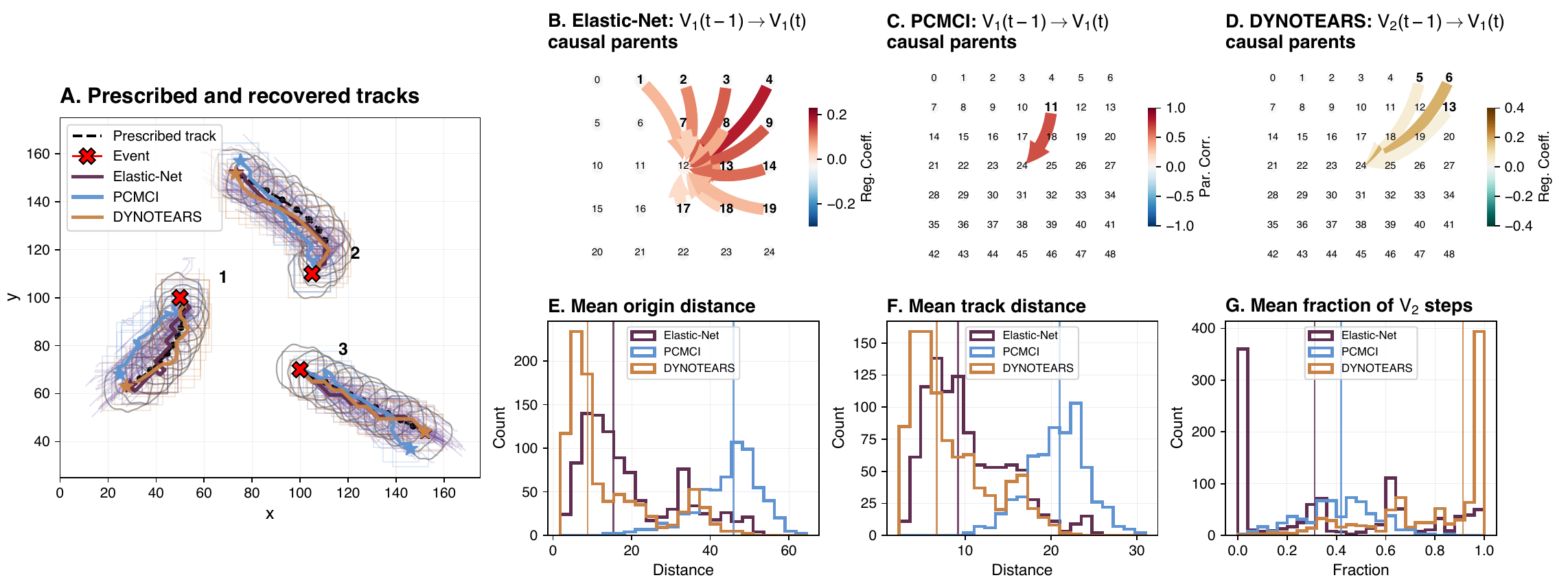}
    \caption{\textbf{TraCE-ST on a two-variable synthetic system with prescribed causal trajectories.}
        (\textbf{A}) Three examples of prescribed and recovered causal trajectories (from $t_0$ to $t_{-30}$). Red $\boldsymbol{\times}$ markers denote target children at $t_0$, and gray contours show prescribed Gaussian perturbations in $V_1$ (anomaly $>0.5$ over $t_0$––$t_{-30}$). TraCE-ST trajectories were computed using Elastic-Net, PCMCI, and DYNOTEARS across 1000 random hyperparameter combinations (see Materials and Methods), with 993, 597, and 935 successful runs, respectively. Colored trajectories and boxes denote the best tracks for each of the three ensembles, based on distance to the Gaussian perturbation at $t_{-30}$, while thin purple lines show 100 randomly selected Elastic-Net trajectories.
        (\textbf{B--D}) Causal stencils $\mathcal{B}_r(s)$ for the first TraCE-ST iteration of Track 2 from the best tracks shown in (\textbf{A}) using the three methods ($r=2$, $r=3$, and $r=3$, respectively). Colors indicate causal strength, and bold numbers mark nodes in the selected cluster. Note that, for DYNOTEARS, instead of showing $V_1(t-1)\to V_1(t)$ relationships, we show $V_2(t-1)\to V_1(t)$ (see different color scale), since this method often incorrectly identifies $V_2(t-1)$ as the winning causal parent of $V_1(t)$ (see Supp. Fig. \ref{fig:sup_dyno}).
        (\textbf{E--G}) Distributions of (i) distance between trajectory endpoints and the prescribed track at $t_{-30}$, (ii) pointwise trajectory distance, and (iii) fraction of total trajectory steps within an ensemble incorrectly assigning $V_2$ as a winning causal parent (lower values indicate better performance). Vertical lines show the mean values for each distribution.}
	\label{fig:synth1}
\end{figure}

Here, $V_1$ is causally defined by its own past states, while 
\begin{equation}
V_2(t) =
\phi\,V_2(t-1) +
\sigma\,V_1(t-1) +
\nu
\label{eq:toy_case1}
\end{equation}
where $\phi=0.8$ and $\sigma=0.8$ specify that $V_2(t)$ depends on both itself and $V_1(t-1)$. Therefore, $V_2(t-1)$ does not cause $V_1(t)$, but is correlated with it. We apply TraCE-ST to the three examples of Gaussian perturbations with prescribed trajectories shown in Figure \ref{fig:synth1}A, whose causal origins should be traceable exclusively within $V_1$. We evaluated three different underlying causal discovery engines: DYNOTEARS \cite{pamfil_dynotears_2020}, an Elastic-Net-based Granger-causal discovery procedure \cite{zou_regularization_2005}, and PCMCI \cite{runge_detecting_2019}. The Elastic-Net-based procedure falls within the framework of Granger causality \cite{shojaie_granger_2022}, in which a time series $x_1$ is considered Granger-causal for another series $x_2$ if incorporating past values of $x_1$ in a predictive model improves the prediction of $x_2$, typically conditional on the past of $x_2$ and other relevant variables. DYNOTEARS is closely related, as it estimates lagged dependencies in time-series structure learning, but it is more precisely a score-based dynamic Bayesian network method that jointly estimates the full directed graph across all variables simultaneously, rather than fitting one child variable at a time as in Elastic-Net regression. PCMCI follows a constraint-based approach, where rather than relying on prediction error alone, it uses iterative conditional-independence testing to determine whether dependencies persist after conditioning on selected past variables and other relevant time-series parents. 

We investigated TraCE-ST behavior and characterized its sensitivity to hyperparameter choices by running 1000 trajectories with different hyperparameters for each of the three underlying causal discovery engines (see Materials and Methods), and only evaluated ``successful runs,'' which we defined as those trajectories that ran during the required 30 time steps ($N_{\mathrm{samples}}=935, 993,$ and $597$ for DYNOTEARS, Elastic-Net, and PCMCI, respectively)—omitting those that stopped early due to not finding any causal parent at a given tracking step. As observed in Figure \ref{fig:synth1}A, TraCE-ST can successfully recover the prescribed propagation pathways using any of the algorithms, and, for suitable hyperparameter choices within the explored ranges, any of them can achieve a relatively low trajectory reconstruction error and accurately locate the Gaussian perturbations 30 time steps back in time (Figure \ref{fig:synth1}E, F). However, the hyperparameter exploration revealed performance differences among the methods (Fig. \ref{fig:synth1}E--G). DYNOTEARS shows the lowest average distance to the prescribed tracks, but often incorrectly selects $V_2$ as the main causal parent (Fig. \ref{fig:synth1}D--G). On the other hand, PCMCI remains more robust to spurious parent selection but, in addition to having higher sensitivity to hyperparameter choices, exhibits less accurate tracks on average and has only a 59.7$\%$ success ratio (597/1000; Fig. \ref{fig:synth1}E--G). Finally, Elastic-Net provides the most stable performance across hyperparameter configurations, consistently remaining close to the prescribed tracks and correctly identifying $V_1$ as the main causal parent in most cases (Fig. \ref{fig:synth1}A, E--G), while being substantially lower in computational cost than DYNOTEARS and PCMCI. These results support the conclusion that TraCE-ST can recover the correct propagation pathway of a perturbation to a target child in the presence of a correlated non-causal variable, although reconstruction performance depends on the hyperparameter choices explored and the underlying causal discovery algorithm.

\subsubsection*{Causal reconstruction in a three-variable system}

\begin{figure}[t!]
	\centering
	\includegraphics[width=1\textwidth]{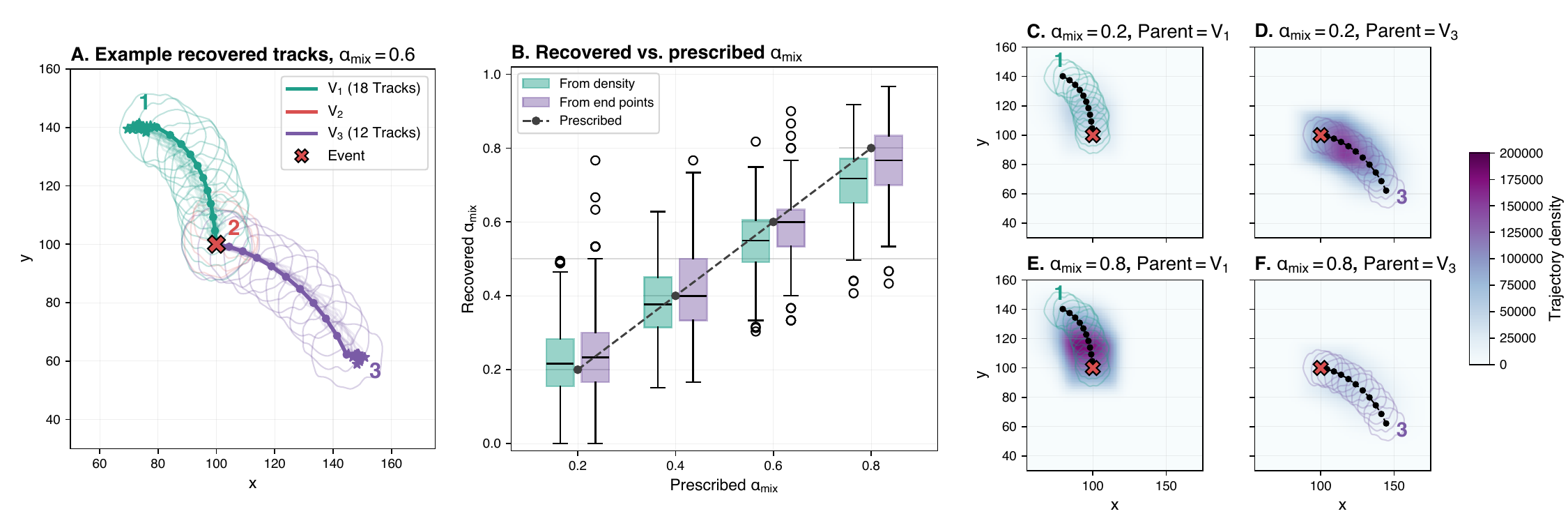}
    \caption{\textbf{TraCE-ST on a three-variable synthetic system with prescribed causal trajectories.}
        (\textbf{A}) Prescribed perturbations and example of recovered trajectories for the three-variable synthetic case ($M=30$, $\alpha_{mix}=0.6$). Independent perturbations propagate in $V_1$ and $V_3$ before interacting to produce the target child in $V_2$. (\textbf{B}) Boxplots of recovered versus prescribed $\alpha_{mix}$, based on trajectory density and endpoint fractions for $V_1$, across 300 30-member ensembles using Elastic-Net. Equivalent results are found using $V_3$ (see Fig. \ref{fig:sup_alpha}). (\textbf{C--F}) Spatial trajectory density across ensembles for $V_1$ and $V_3$ as causal parents at $\alpha_{mix}=0.2$ and $\alpha_{mix}=0.8$. Higher trajectory density indicates that the corresponding variable and location are more frequently identified as causal parents.}
	\label{fig:synth2}
\end{figure}

This case illustrates how the probabilistic formulation of TraCE-ST can estimate the relative contribution of each variable toward causing the target child at $t=0$. In this experiment, over the last eight time steps, the target child $V_2$ arises from a lagged mixture of prescribed perturbations propagating independently in $V_1$ and $V_3$ that collide spatially. Specifically,
\begin{equation}
V_2(t) =
\alpha_{mix}\,V_1(t-1) +
(1-\alpha_{mix})\,V_3(t-1) +
\nu
\label{eq:toy_case2}
\end{equation}
for $t \in \{t_0 - 7, \dots, t_0\}$, where $\alpha_{mix}$ specifies the relative contribution of $V_1$ as a causal parent to the target child, whereas $1-\alpha_{mix}$ specifies the relative contribution of $V_3$. As in the previous case, and given its better performance and efficiency, we performed a broad hyperparameter exploration using Elastic-Net. We ran TraCE-ST trajectories with $\alpha_{mix}=0.2, 0.4, 0.6, 0.8$ and 300 different hyperparameter combinations, each yielding an ensemble size $M=30$. As shown in Figure \ref{fig:synth2}A,C--F, the probabilistic version of TraCE-ST can correctly reconstruct the two families of co-occurring prescribed pathways, with trajectories remaining close to one of the prescribed tracks. Additionally, the normalized trajectory density, as well as the fraction of trajectories with $V_1$ as causal parent at $t_{-30}$, are approximately linearly related to the prescribed $\alpha_{mix}$ (Fig. \ref{fig:synth2}B--F). Together, our synthetic experiments demonstrate that TraCE-ST can recover causal trajectories and correctly estimate the relative contributions of multiple parents causing a target child, providing confidence in its application to more realistic Earth system phenomena.

\subsubsection*{Benchmarking Underlying Causal Discovery Engines}

To further evaluate our choice of Elastic-Net over PCMCI and DYNOTEARS for the real-world TraCE-ST applications that follow, we performed two additional tests with synthetic cases. First, we ran a single-trajectory computational timing benchmark using the three causal discovery algorithms with consistent TraCE-ST hyperparameters, varying only underlying-algorithm-specific parameters. This benchmark fairly evaluates computational timing as a determining factor for choosing Elastic-Net instead of PCMCI or DYNOTEARS. PCMCI requires approximately 3–4 times more wall-clock time than Elastic-Net, while DYNOTEARS requires nearly two orders of magnitude more time (Elastic-Net: 3.35 s; PCMCI: 13.8 s; DYNOTEARS: 300.4 s per trajectory). Combined with its higher incorrect-parent attribution rate in the two-variable synthetic case, the substantially higher computational cost of DYNOTEARS makes it less suitable for the larger ensemble, real-world applications that follow. 

Second, we used TraCE-ST with PCMCI as the underlying causal engine for the three-variable synthetic case to compare attribution accuracy with results from the Elastic-Net-based three-variable analysis above. We used 30 hyperparameter configurations and 20-member ensembles for each prescribed $\alpha_{mix}$, and consistent TraCE-ST hyperparameters from the Elastic-Net-based three-variable synthetic case. PCMCI-specific hyperparameters followed the PCMCI ranges used in the two-variable synthetic case (see Materials and Methods). Consistent with the results from the two-variable synthetic case, PCMCI showed lower success in recovering the prescribed relative contribution of $V_1$ and $V_3$ as causal parents (Fig.~\ref{fig:sup_alpha_pcmci}). 

These benchmarking results indicate that TraCE-ST can be used with various causal discovery engines, but that quantitative attribution accuracy and computational feasibility can depend strongly on the selected engine. Together, trajectory accuracy, robustness to hyperparameter choices, and computational efficiency motivate the adoption of Elastic-Net as the causal discovery engine in the following real-world applications.

\subsection*{Three real-world scenarios}

We now evaluate TraCE-ST in three real-world cases where the dominant causal pathways are sufficiently well understood to enable concise validation: (i) Tropical Storm Debby (2006), a fast-evolving tropical system influenced by remote vorticity disturbances; (ii) the Mount Pinatubo eruption (1991), which injected large amounts of gases into the stratosphere, where they were transported circumglobally, formed aerosols, and drove significant radiative impacts; and (iii) the 2021 Pacific Northwest (PNW) heatwave, a deadly, extreme event produced by a mid-tropospheric ridge arising from the combined influence of local and remote drivers. Together, these cases span distinct components of the Earth system, a wide range of spatiotemporal scales, varying levels of dimensional complexity, and multiple datasets and geographic areas of interest. For each case, we explore hyperparameter combinations to encourage spatially coherent trajectory evolution and ensembles that satisfy predefined trajectory-completion criteria. Additionally, variables are standardized to place different physical quantities on comparable numerical scales before causal discovery, although other preprocessing choices may be preferable for highly skewed or intermittent variables \cite{furtado_setting_2026}.

\subsubsection*{(i) Tropical Storm Debby (2006)}

Debby developed over the eastern North Atlantic Ocean from an African easterly wave that moved off the west coast of Africa on 20 August 2006, becoming a tropical depression by 1800 UTC on 21 August \cite{franklin2006debby}. Debby subsequently intensified into a tropical storm at 0000 UTC on 23 August while moving northwestward over the Atlantic. Beyond a favorable large-scale environment, several studies have shown that Debby's genesis and evolution were strongly influenced by the westward propagation, merging, and organization of upstream vorticity disturbances over the African continent during the 10 days prior to achieving tropical depression status \cite{sippel_environmental_2011,lin_origin_2013,chen_relation_2014,zhu_genesis_2015}. These disturbances included both dry and wet vortices, the latter promoting organized deep convection, including mesoscale convective systems embedded within an African easterly wave, whose cold cloud tops are reflected in brightness temperature observations. The evolution of these vorticity disturbances culminated in a dry-wet vortex merger near the West African coast and contributed to the development of Debby \cite{chen_relation_2014}. Using precipitation occurring within the storm as the target child $C^{(j_0)}(\mathcal{R}_0,t_0)$, we assess whether TraCE-ST can recover Debby's track and identify upstream vorticity disturbances over Africa as the causal parents with the highest integrated trajectory density throughout the trajectories, which would be consistent with the aforementioned studies. Additionally, we include brightness temperature in the analysis to represent convective cloudiness and mesoscale convective systems, which are also causal parents of the targeted precipitation.

Using hourly data on a $0.25^\circ \times 0.25^\circ$ grid, we apply TraCE-ST to three variables: IMERG precipitation (i.e., rain rate; mm h$^{-1}$) \cite{huffman_integrated_2020}, brightness temperature (K) from NCEP/CPC as a proxy for cloudiness \cite{janowiak2017ncep}, including mesoscale convective systems, and 500--700-hPa vorticity (s$^{-1}$) from ERA5 \cite{hersbach_era5_2020}. IMERG is preferred over ERA5 precipitation because its heavier reliance on satellite retrievals generally improves the representation of precipitation variability at subdaily timescales \cite{tang_have_2020}, and brightness temperature from NCEP/CPC is preferred over ERA5-derived cloud variables because ERA5 can underestimate key aspects of mesoscale convective activity \cite{yu_mesoscale_2025}. We acknowledge that other variables beyond precipitation could represent Debby, such as surface pressure or wind anomalies, but we choose precipitation because it can be measured more accurately at hourly timescales and is a consequence of the storm's presence. Additional variables could also be considered causal parents, including large-scale environmental fields such as SSTs and wind shear, but we focus here on the demonstrative use of TraCE-ST in a case study supported by current literature \cite{sippel_environmental_2011,lin_origin_2013,chen_relation_2014,zhu_genesis_2015} and leave analysis of other potential causal parents of Debby for future work. Standardized hourly anomalies for precipitation, cloudiness, and mid-level vorticity were computed by removing the hourly mean and dividing by the hourly standard deviation of each grid cell for the 2006 July-to-September period, and trajectories are initialized at 00:00 UTC on 26 August 2006 (around 4 days after Debby became a tropical depression) at a target location near Debby's late-stage tropical-storm circulation coinciding with an area of high precipitation and low brightness temperature according to IMERG and NCEP/CPC (28$^\circ$N, 46$^\circ$W). We generated 300 30-member ensembles of 15-day backward TraCE-ST trajectories using a region $\mathcal{R}$ of size 5--7$^\circ \times$5--7$^\circ$ and a temporal window $\mathcal{W}$ of 5--7 h (see Materials and Methods). Of the 300 ensembles, we retained 68 for analysis. Each retained ensemble had a mean trajectory length exceeding 12 days, with fewer than 10 ensemble members producing trajectories shorter than 12 days. These criteria were used to identify hyperparameter settings that produced sufficiently persistent trajectories for the respective case study.

\begin{figure}[t!]
	\centering
	\includegraphics[width=1\textwidth]{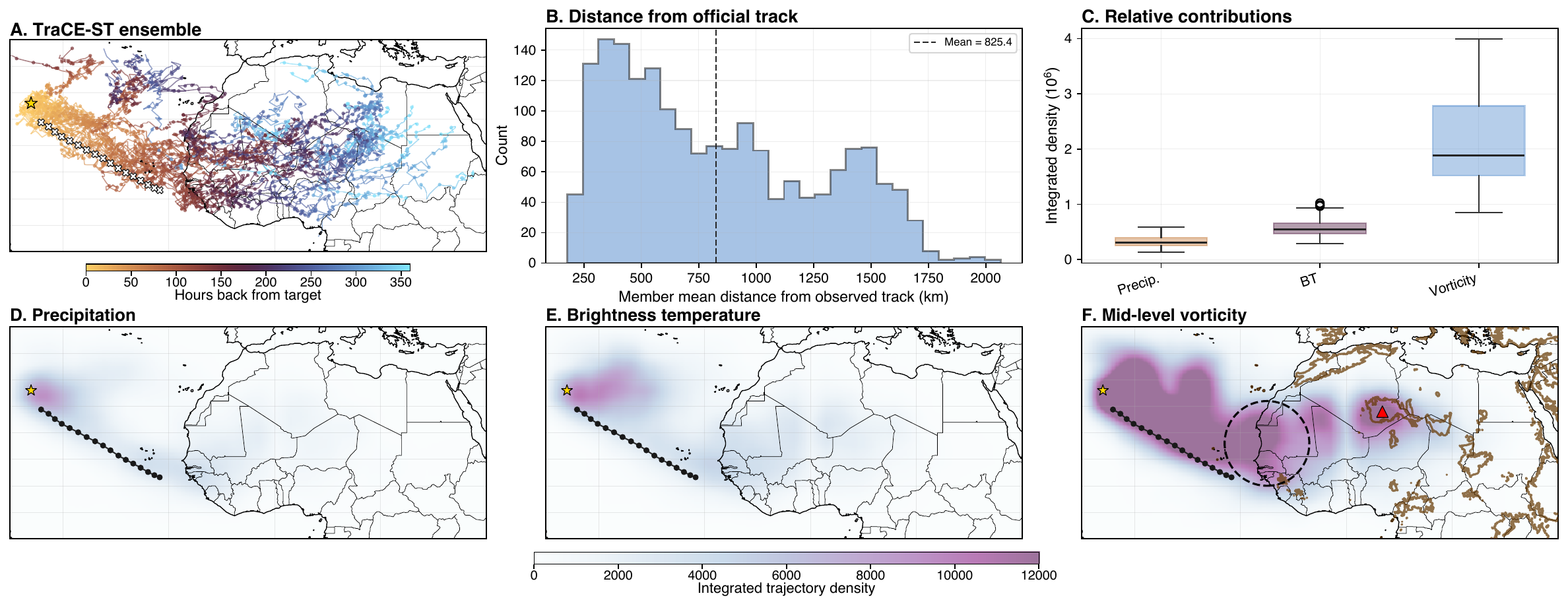}
	\caption{\textbf{TraCE-ST on Tropical Storm Debby (2006).}
        (\textbf{A}) Example ensemble of TraCE-ST trajectories initialized at 00:00 UTC on 26 August 2006 at a target location of Debby's late-stage tropical-storm circulation (28$^\circ$N, 46$^\circ$W; yellow star), using brightness temperature, mid-level vorticity, and precipitation (target child at $t=0$). Colors indicate hours backward from the target time, and $\boldsymbol{\times}$ markers denote the 6-hourly Debby's track from IBTrACS from 21 August 1800 UTC to 26 August 0000 UTC \cite{knapp_international_2010}. 
        (\textbf{B}) Distribution of the mean pointwise distance between TraCE-ST trajectories and the IBTrACS track across ensembles (for the 21 August 1800 UTC--26 August 0000 UTC period). The mean pointwise distance is 825.4km, and, for reference, the total length of the plotted Debby's track ($\boldsymbol{\times}$ markers in A) is 2769.6km, while the average distance per 6-hourly time step is 173.1km.
        (\textbf{C}) Boxplots of estimated relative contributions (integrated trajectory density across ensembles) between 5 and 15 days prior to the target child for each variable. 
        (\textbf{D--F}) Variable-specific trajectory densities for precipitation, brightness temperature, and mid-level vorticity, respectively, across ensembles; darker colors indicate higher densities and thus greater inferred causal contributions. The observed Debby track from IBTrACS is also overlaid in black for reference. Brown contours in F highlight the topography at 800 meters above sea level, the red triangle points at the Hoggar Mountains, and the dashed circle highlights the location of the dry-wet vortex merger according to Chen and Liu \cite{chen_relation_2014}.}
	\label{fig:debby}
\end{figure}

Most TraCE-ST trajectories align with Debby's observed track for several days before the target child, with brightness temperature, precipitation, and mid-level vorticity all contributing as causal parents. A relatively small but considerable fraction of trajectories diverges early from the observed storm track into filaments of vorticity, and less frequently brightness temperature (Fig. \ref{fig:debby}A,E,F). This behavior is physically consistent with tropical storm organization and intensification, in which convergent flow and convective activity can concentrate cyclonic vorticity near the storm core, supporting inner-core spin-up and heavy precipitation \cite{smith_toward_2015}. 

About 5--8 days back, the highest trajectory densities concentrate near the West African coast and the adjacent eastern Atlantic, where Debby's pre-depression circulation organized (Fig. \ref{fig:debby}F). One of the main mechanisms for Debby's formation took place over these areas, as identified by Chen and Liu \cite{chen_relation_2014}: a dry vortex coming from North Africa and passing through Senegal merged with a southeasterly wet vortex passing through Guinea, producing a high-rotation core that ultimately formed the tropical storm over the ocean. This phenomenon is clear in Figure \ref{fig:debby}A,F (see contoured area in \ref{fig:debby}F, where high density of trajectories tends to concentrate), highlighting TraCE-ST's ability to represent the multiple causal parents driving Debby's formation. After spreading over continental Africa, trajectories extend inland over Algeria, Libya, Niger, and Chad. The backward TraCE-ST trajectories do not collapse onto a single narrow filament. Instead, they reveal a diffuse upstream structure that extends eastward across the Sahel between 8 and 15 days before the target child's date (Fig.~\ref{fig:debby}A,F). High trajectory density is located over the Hoggar Mountains, south of Algeria (see topography in Fig. \ref{fig:debby}F), which have been identified to favor vortex genesis during July and August, contributing in some cases to cyclogenesis \cite{duvel_vortices_2021,duvel_origin_2023}. The obtained spatial distribution of trajectory density is consistent with the disorganized, propagating vorticity disturbances coming from both mesoscale convective systems and drier northern regions that ultimately feed the dry-wet vortex merger \cite{lin_origin_2013,zhu_genesis_2015}. 

Overall, 5--15 days before the target child, trajectories span a broader precursor region over western Africa (Figs. \ref{fig:debby}A and \ref{fig:sup_ens_debby}), during which time, trajectory densities estimate that mid-level vorticity is more frequently a causal parent of the target child than brightness temperature and precipitation, consistent with past studies \cite{sippel_environmental_2011,lin_origin_2013,chen_relation_2014,zhu_genesis_2015} (Figs. \ref{fig:debby}C and \ref{fig:sup_dens_debby}). Taken together, these results indicate that TraCE-ST not only reproduces the observed storm track but also identifies physically consistent causal trajectories. The highest trajectory density is associated with mid-level vorticity organization over West Africa, while upstream mid-level vorticity disturbances over the more central part of the continent (mainly over the Hoggar Mountains) also fed Debby's emergence.

\subsubsection*{(ii) Mount Pinatubo eruption (1991)} 

The 15 June 1991 eruption of Mount Pinatubo (15$^\circ$N, 121$^\circ$E) produced one of the most consequential stratospheric aerosol perturbations of the twentieth century \cite{mccormick_atmospheric_1995}, with impacts on radiative fluxes, surface and stratospheric temperatures, and the hydrological cycle \cite{hansen_potential_1992,brown_validating_2024}. The broad physical pathway is well established: large amounts of sulfur dioxide gas (SO$_2$) were injected into the stratosphere and oxidized to sulfuric acid (H$_2$SO$_4$), which then nucleated or condensed to form sulfate aerosol particles, represented here by SO$_4$ burden. The transport and growth of these sulfate aerosols dominated the evolving aerosol optical depth \cite{mccormick_atmospheric_1995,brown_validating_2024}. This case, therefore, offers a particularly useful validation target for TraCE-ST because the event of interest involves not only tracer transport but also a known sequence of cross-variable transformations.

\begin{figure}[t!]
	\centering
	\includegraphics[width=1\textwidth]{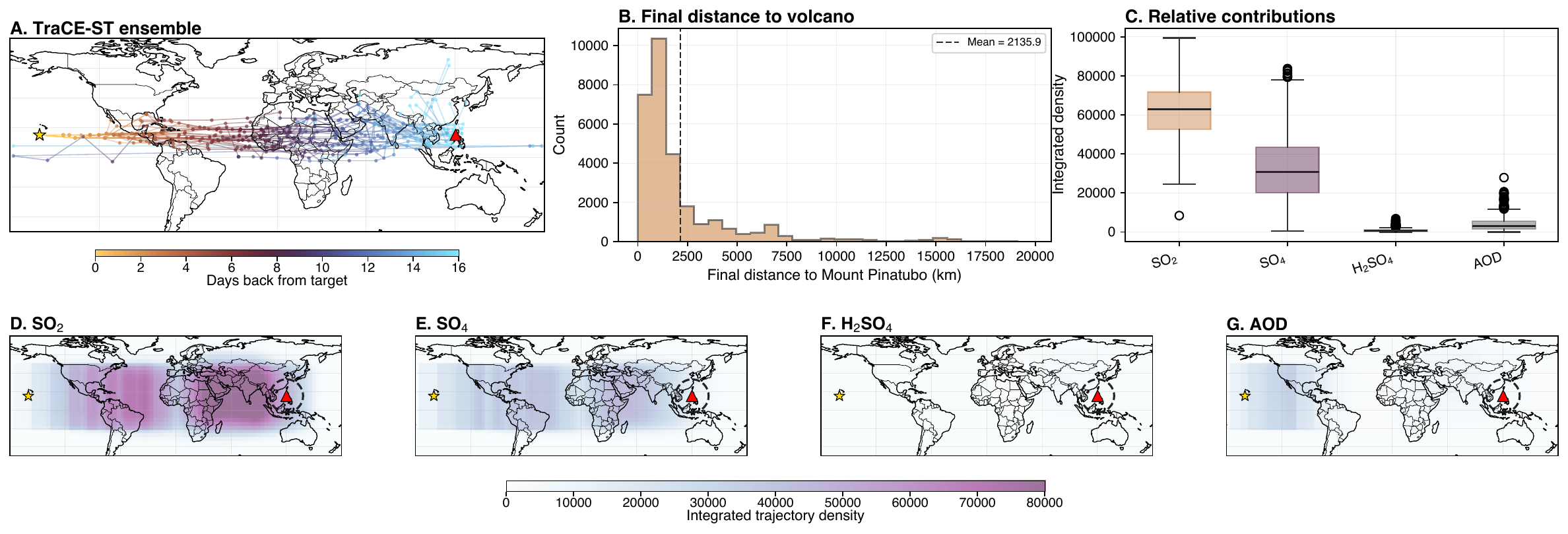}
	\caption{\textbf{TraCE-ST on Mount Pinatubo eruption (1991).}
        (\textbf{A}) Example ensemble of TraCE-ST trajectories initialized on 2 July 1991 (17 days after Mount Pinatubo eruption) over the central Pacific (15$^\circ$N, 160$^\circ$W; yellow star), using SO$_2$, H$_2$SO$_4$, and SO$_4$ burden, together with visible aerosol optical depth (AOD) as the target child at $t=0$, from an E3SMv2-SPA simulation \cite{brown_validating_2024}. Mount Pinatubo's location is indicated by the red triangle.
        (\textbf{B}) Distribution of distances between Mount Pinatubo and trajectory endpoints across ensembles ($t-16\,\mathrm{d}$). 
        (\textbf{C}) Boxplots of estimated relative contributions, defined as integrated trajectory density across ensembles, between 4 and 16 days prior to the target child for each variable. 
        (\textbf{D--G}) Variable-specific trajectory densities across the ensembles. As a distance reference, a dashed circle with a radius of 2000 km is plotted around the location of Mount Pinatubo.}
	\label{fig:pinatubo}
\end{figure}

We apply TraCE-ST to daily data at a 4$^\circ \times$4$^\circ$ grid spacing from the control E3SMv2-SPA simulation of Brown et al. \cite{brown_validating_2024}, which reasonably reproduces the post-Pinatubo evolution of sulfur burden and stratospheric aerosol. We use four variables: SO$_2$ burden, H$_2$SO$_4$ burden, SO$_4$ burden, and visible aerosol optical depth, with the latter used as the target child at $t=0$. Each variable was standardized separately at each grid cell by subtracting its 1991 mean and dividing by its 1991 standard deviation within the simulation. Trajectories were initialized over the central Pacific (15$^\circ$N, 160$^\circ$W) on 2 July 1991 (17 days after the eruption), when elevated aerosol optical depth was already evident far from the volcano, and were integrated backward at a daily time step for 16 days, which is sufficient time for chemical reactions and long-range transport to take place. We generated 1000 ensembles with 30 members each using a region $\mathcal{R}$ of size 65--75$^\circ \times$65--75$^\circ$ and a temporal window $\mathcal{W}$ of 3--5 d (see Materials and Methods). All 1000 ensembles were retained because more than 20 of their 30 trajectories (members) successfully ran during 16 days. We test whether TraCE-ST trajectories converge toward the eruption region as they transition from optical-depth-dominated to sulfur-dioxide-dominated states.

As trajectories from the TraCE-ST ensembles are integrated back in time, they converge toward the eruption location, being within approximately 2000 km of the volcano at $t-16\,\mathrm{d}$ (Figs. \ref{fig:pinatubo}A,B, and \ref{fig:sup_ens_pinatubo}). TraCE-ST also identifies SO$_2$ as the variable with comparatively higher trajectory density (as a causal parent) 4-to-16 days before July 2, the date associated with the target child (Fig. \ref{fig:pinatubo}C). The framework further recovers the expected cross-variable behavior (Fig. \ref{fig:pinatubo}D--G): near the target child's time over the Pacific (1--3 d), sulfate aerosol burden (SO$_4$) is more frequently found as a causal parent than SO$_2$ (Fig. \ref{fig:sup_dens_pinatubo}), whereas later along the trajectory, SO$_4$ trajectory density is present but lower than that of SO$_2$, consistent with SO$_4$'s role as a causal parent of aerosol optical depth and a causal child of SO$_2$ (Fig. \ref{fig:pinatubo}C,E). Across the ensemble, trajectories increasingly converge toward SO$_2$ as a causal parent and approach the volcanic source region (Fig. \ref{fig:pinatubo}D), consistent with the known sulfur injection and subsequent circumglobal transport after the eruption \cite{mccormick_atmospheric_1995}. In contrast, H$_2$SO$_4$ does not emerge frequently as a causal parent along TraCE-ST trajectories (Fig.~\ref{fig:pinatubo}C,F). This result is consistent with H$_2$SO$_4$ acting as a short-lived intermediate whose production and condensation may occur on timescales shorter than the daily data resolution, limiting TraCE-ST's ability to resolve this parent directly. These results show that, even when some chemical and microphysical transformations are under-resolved, TraCE-ST can recover the expected large-scale transport pathway and post-eruption gas-to-aerosol transformation.

\subsubsection*{(iii) 2021 North American Pacific Northwest heatwave} 

In late June 2021, the Pacific Northwest (PNW) of North America experienced an unprecedented heatwave, with many locations breaking all-time maximum 2-meter ambient temperature records by more than 5°C and causing severe societal and ecological impacts \cite{white_unprecedented_2023,schiermeier2021climate,reyes_high-elevation_2023,mass_pacific_2024}. The event's exceptional magnitude has been attributed to interacting processes across variables and regions, including a persistent blocking ridge, land-atmosphere feedbacks, upstream diabatic heating and moisture transport, a split polar-vortex configuration, and comparatively less explored ocean-atmosphere heat exchanges \cite{neal_2021_2022,schumacher_drivers_2022,li_role_2024,oertel_everything_2023,baier_3-week-long_2023,overland_causes_2021,wang_unprecedented_2023,duan_impact_2025}. The blocking ridge promoted subsidence and adiabatic warming while reducing cloud cover and thereby increasing incoming shortwave radiation over the PNW \cite{neal_2021_2022,schumacher_drivers_2022,li_role_2024}. Land-atmosphere interactions further amplified the event, as pre-existing soil-moisture deficits altered latent and sensible heat fluxes and shifted the surface energy balance toward sensible heating. Identifying these factors and quantifying their relative importance has proven challenging, motivating multiple methodological approaches, including numerical modeling, reanalysis diagnostics, and air parcel tracking \cite{mo_anomalous_2022,oertel_everything_2023,baier_3-week-long_2023,neal_2021_2022,mckinnon_how_2022,lin_2021_2022,duan_impact_2025,schumacher_drivers_2022,fleishman_synthesis_2025,overland_causes_2021,li_role_2024}. Here, we test whether TraCE-ST can identify these processes in a data-driven manner by identifying the spatiotemporal pathways through which different variables contributed to the development and amplification of the blocking ridge.

\begin{figure}[t!]
	\centering
	\includegraphics[width=1\textwidth]{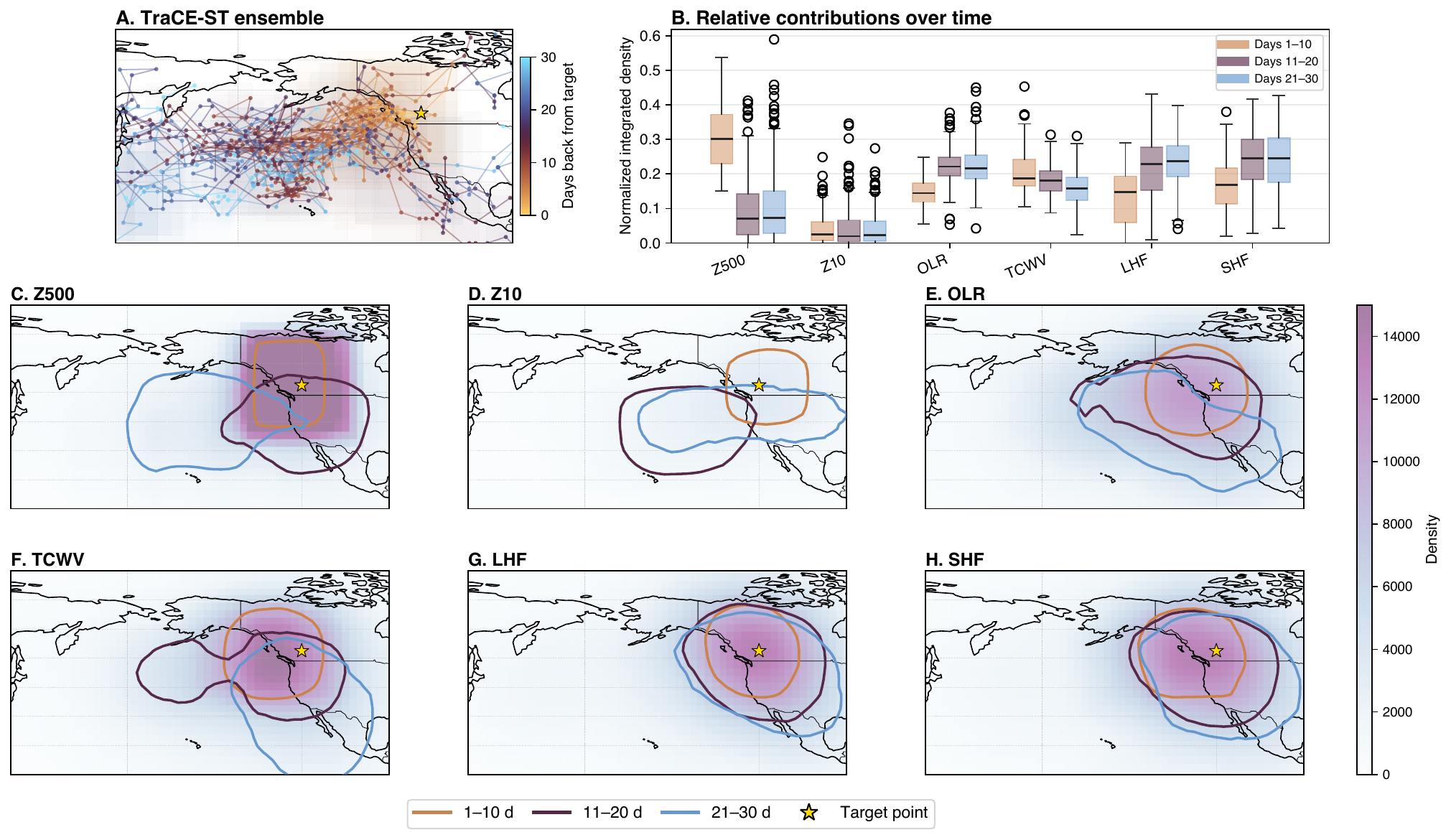}
	\caption{\textbf{TraCE-ST on the 2021 PNW heatwave.}
        (\textbf{A}) Example ensemble of TraCE-ST trajectories. Colors indicate the number of days before the target date (June 30, 2021), and the yellow star shows the location of the target child (52.5°N, 120°W). Each trajectory was integrated for up to 30 days using daily ERA5 data at 2° grid spacing. (\textbf{B}) Normalized spatially integrated trajectory density across ensembles stratified by variable for three lead-time windows: 1--10, 11--20, and 21--30 days before June 30. Higher values indicate greater relative causal frequency. (\textbf{C--H}) Spatial trajectory density for each variable across ensembles. Colored contours indicate the 95th percentile of density for each variable during three time windows, highlighting regions of highest causal frequency.}
	\label{fig:pnw1}
\end{figure}

To trace the causes of the intense blocking ridge and associated heatwave, we defined the target child as the 500-hPa geopotential (Z500) anomaly on June 30, 2021, over British Columbia (centered at 52.5°N, 120°W), and TraCE-ST trajectories were integrated backward for 30 days. To represent the processes discussed above, we included six daily anomaly variables from ERA5 at 2° grid spacing: Z500, 10-hPa geopotential (Z10), outgoing long-wave radiation (OLR), total column water vapor (TCWV), mean surface latent heat flux (LHF), and mean surface sensible heat flux (SHF). Table \ref{tab:pnw_variables} connects each variable to the processes described above. For each grid cell, anomalies relative to the 1995-2024 climatology were computed by subtracting the daily climatological mean and dividing by the daily climatological standard deviation, with both quantities smoothed using a 30-day window. We used a region $\mathcal{R}$ of size 30--40$^\circ \times$30--40$^\circ$ and a temporal window $\mathcal{W}$ of 4--6 d (see Materials and Methods), and randomly sampled 300 30-member ensembles with different hyperparameter combinations. Of the 300 ensembles, 242 were retained because their trajectories were successfully traced for 25 out of 30 time steps on average. Consistent with the literature, most trajectories originate west of the target child, although this real-world case study shows greater sensitivity to hyperparameters than the two previous cases (Figs. \ref{fig:pnw1}A and \ref{fig:sup_ens_pnw21}). 

\begin{table}[t!]
	\centering
	\caption{\textbf{Variable anomalies used in the PNW heatwave analysis.}
	Summary of the main physical processes represented by each variable included in the TraCE-ST analysis.}
	\label{tab:pnw_variables}
	
	\begin{tabular}{lp{11.2cm}}
		\\
		\hline
		Variable & Potential causal drivers being represented\\
		\hline
		Z500 (m$^2$ s$^{-2}$) & Mid-tropospheric circulation and Rossby-wave dynamics \cite{schumacher_drivers_2022,neal_2021_2022}.\\
		Z10 (m$^2$ s$^{-2}$) & Stratospheric precursors, including polar-vortex variability \cite{overland_causes_2021,wang_unprecedented_2023}.\\
		OLR (W m$^{-2}$) & Cloudiness, deep convection, and diabatic forcing over the tropical and subtropical Pacific \cite{oertel_everything_2023,baier_3-week-long_2023}.\\
		TCWV (kg m$^{-2}$) & Moisture availability associated with warm conveyor belts or atmospheric river moisture transport \cite{neal_2021_2022,mass_pacific_2024,oertel_everything_2023,white_unprecedented_2023}.\\
		LHF (W m$^{-2}$) & Ocean--atmosphere and land--atmosphere coupling through latent heat exchange \cite{schumacher_drivers_2022,duan_impact_2025}.\\
		SHF (W m$^{-2}$) & Ocean--atmosphere and land--atmosphere coupling through sensible heating \cite{schumacher_drivers_2022,duan_impact_2025}.\\
		\hline
	\end{tabular}
\end{table}

\begin{figure}[t!]
	\centering
	\includegraphics[width=1\textwidth]{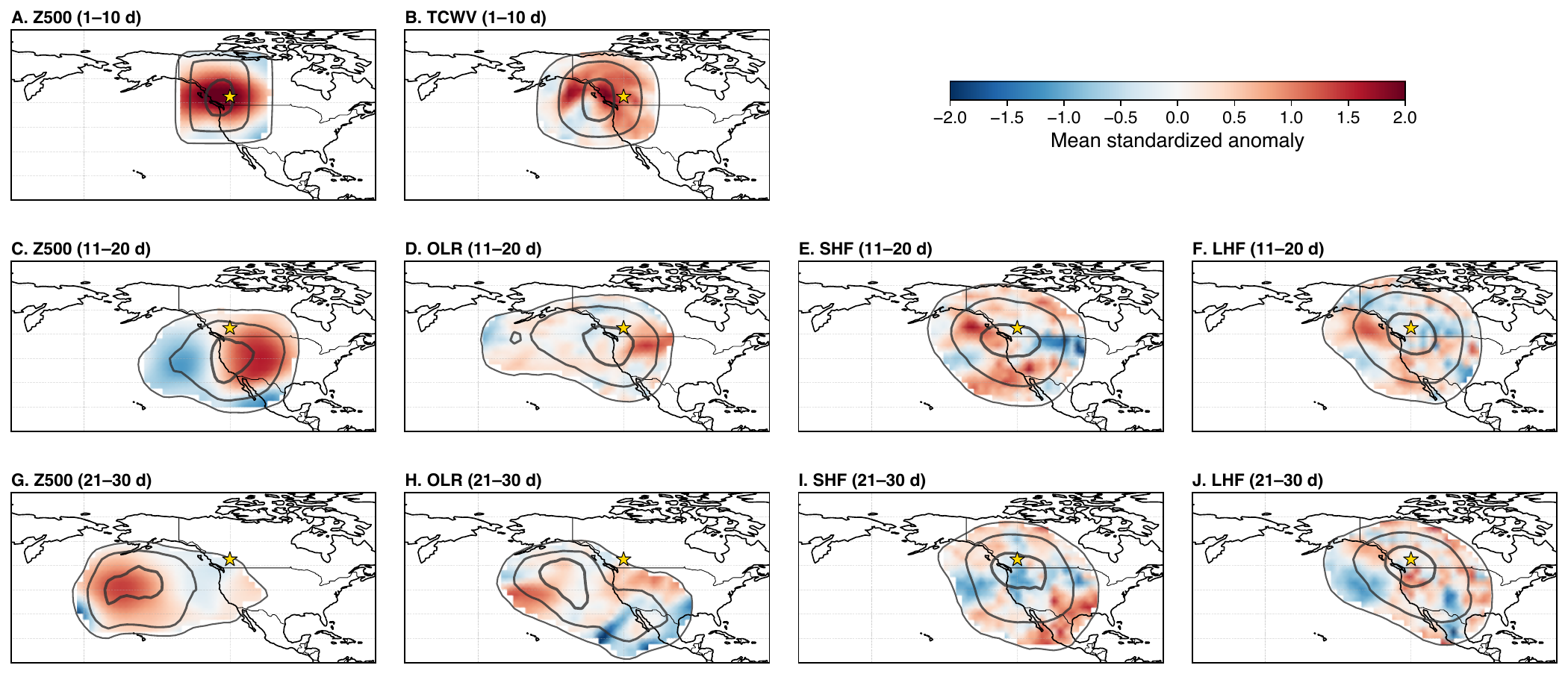}
	\caption{\textbf{Relevant anomaly patterns identified by TraCE-ST for the 2021 PNW heatwave.}
        (\textbf{A--J}) Colored shading shows the mean standardized daily anomalies of the most frequent causal parents during the three lead-time windows identified in Fig.~\ref{fig:pnw1}: 1--10 days before the target child (A,B), 11--20 days before the target child (C--F), and 21--30 days before the target child (G--J). Black contours mark regions where TraCE-ST density exceeds the 90th, 95th, and 99th percentiles for the corresponding variable and window, highlighting the locations that repeatedly emerge as causal parents. Note that Z10 results are not included in this plot, given the low trajectory density in Figure \ref{fig:pnw1}.}
	\label{fig:pnw2}
\end{figure}

During the 1--10 days before the heatwave peak (June 20--29), Z500 and TCWV emerge as the more frequent causal parents, with trajectory density maxima centered immediately upstream of the target child at $t=0$ (Fig. \ref{fig:pnw1}B,C,F; see orange contours). These results are consistent with the established mechanism for ridge amplification, in which the block intensifies as it propagates from ocean to land, while moisture transport associated with an atmospheric-river-like feature enhances latent heat release \cite{schumacher_drivers_2022,neal_2021_2022,oertel_everything_2023,mo_anomalous_2022,white_unprecedented_2023}. During the 11--20 days before the heatwave peak (June 10--19), the frequency of Z500 as a causal parent decreases, TCWV remains frequent, and OLR is selected more frequently as a causal parent according to trajectory density (Fig. \ref{fig:pnw1}B). OLR-related trajectories become increasingly frequent across the North and East Pacific (Fig.~\ref{fig:pnw1}E), suggesting that diabatic processes contributed to cyclogenesis and downstream Z500 amplification. These trajectories do not align with the largest magnitude OLR anomalies (Fig.~\ref{fig:pnw2}D), but instead cluster in regions of high OLR variability off the western United States coast and southwest of Alaska (Fig.~\ref{fig:sup_pnw21_std}D). TraCE-ST thus identifies changes in OLR within these regions, rather than the magnitude of the anomalies alone, as contributing to ridge formation. Surface fluxes, represented by LHF and SHF, also emerge more frequently as causal parents during the 11--20 day window (Fig. \ref{fig:pnw1}B). Their trajectories point to contributions from ocean-atmosphere energy exchange over the northeast Pacific and land-atmosphere energy exchange over western North America in the lead-up to the extreme heatwave (Figs. \ref{fig:pnw1}G,H and \ref{fig:pnw2}E,F).

During the 21--30 days before the heatwave (May 31--June 9), OLR remains a frequent causal parent (Fig.~\ref{fig:pnw1}B), with areas of high trajectory density extending into the subtropical eastern Pacific (Figs. \ref{fig:pnw1}E and \ref{fig:pnw2}H). This spatial result includes a region of low OLR (high cloudiness) offshore of Baja California and a region of reduced cloudiness over the central Pacific (Fig. \ref{fig:pnw2}H), consistent with studies linking the heatwave to remote diabatic forcing and organized cloudiness at subseasonal lead times \cite{oertel_everything_2023,baier_3-week-long_2023}. Moreover, the area of high Z500 trajectory density coincides with positive anomalies over the central Pacific (Figs.~\ref{fig:pnw1}C and \ref{fig:pnw2}G), consistent with persistent wave activity spanning the subtropical and extratropical Pacific. LHF and SHF remain frequent causal parents during this period, mainly over western North America and, more weakly, across the Pacific (Fig.~\ref{fig:pnw1}G,H). Over western North America, regions of enhanced LHF and diminished SHF emerge as causal parents for the formation of the event according to TraCE-ST (Figs. \ref{fig:pnw1}G,H, and \ref{fig:pnw2}I,J), underscoring the role of land-surface preconditioning in contributing to the heatwave \cite{duan_impact_2025}. Additionally, the region of high trajectory density for SHF and LHF as causal parents coincides with areas of high anomaly variability over the eastern Gulf of Alaska (Fig. \ref{fig:sup_pnw21_std}I,J). These results suggest that noticeable changes in ocean-surface fluxes played a larger role than what is currently emphasized in the literature. Finally, the density of trajectories for Z10 was relatively low across lead times, indicating that, according to TraCE-ST, relative contributions from the stratosphere as a causal parent were lower than those of the other variables included in our analysis (Figure \ref{fig:pnw1}B,D).

Overall, Figures \ref{fig:pnw1} and \ref{fig:pnw2} show that TraCE-ST is able to identify the multivariate causal pathways leading up to the 2021 PNW heatwave, which include moisture transport and diabatic processes over the northeast Pacific, preceded by large-scale circulation anomalies, remote convective forcing, and land and ocean surface fluxes. These results support central mechanisms proposed in the literature while providing a data-driven estimate of their relative contributions from distinct causal parents over time \cite{neal_2021_2022,oertel_everything_2023,baier_3-week-long_2023,schumacher_drivers_2022}.

\section*{Discussion}

We introduced TraCE-ST, a framework that identifies event-conditioned multivariate causal pathways in space and time using gridded data. TraCE-ST builds upon recent causal discovery advances that make it possible to estimate directed links between grid cells over short time windows \cite{nichol_space-time_2025,nichol_mcastle_2026} by extending them into Lagrangian trajectories that trace the causal parents of a realized outcome of interest. Our approach provides a probabilistic formulation for empirically estimating the relative contribution of multiple causal parents to a given event of interest, rather than reducing causal explanation to a single pathway. This capability is important because Earth system extremes often emerge from multiple interacting processes across variables, regions, and timescales. Overall, TraCE-ST provides an interpretable bridge between local causal discovery and event-scale process attribution, offering an objective and computationally efficient way to formulate data-driven hypotheses that complement association analyses and targeted numerical experiments.

We validated TraCE-ST across a broad range of spatiotemporal scales, levels of complexity, and datasets. In synthetic controlled experiments, TraCE-ST accurately recovered prescribed spatiotemporal propagation pathways, avoided attributing effects to correlated but non-causal variables, and reproduced the imposed relative contributions of multiple causal parents. Furthermore, three real-world applications on extreme events show that TraCE-ST can recover physically meaningful causal pathways in Earth system settings that differ in scale, dominant physics, and complexity. 

For Tropical Storm Debby, TraCE-ST identified the merging of mid-level vorticity as the main driver of storm development. TraCE-ST traced these vorticity-based trajectories upstream across West and North Africa, consistent with previous work emphasizing the roles of mesoscale convective systems and dry vorticity cores over land \cite{lin_origin_2013,chen_relation_2014,zhu_genesis_2015}. In particular, the trajectories point to vortices produced over the Hoggar Mountains in southern Algeria. For Mount Pinatubo, TraCE-ST recovered the numerically modeled large-scale transport and chemical transformation of volcanic emissions from the eruption region to the central Pacific, consistent with the known post-eruption evolution of sulfur, sulfate, and aerosol optical depth \cite{brown_validating_2024}. Finally, for the Pacific Northwest heatwave of 2021, TraCE-ST traced the development and amplification of the blocking ridge through a multivariate causal sequence involving diabatic processes, moisture transport, wave activity, and surface-atmosphere coupling. TraCE-ST trajectories largely agree with mechanisms proposed in previous studies, while also estimating the relative contributions of distinct causal parents and highlighting less studied processes, including the possible role of ocean-atmosphere fluxes over the Gulf of Alaska at subseasonal lead times \cite{schumacher_drivers_2022,neal_2021_2022,oertel_everything_2023,baier_3-week-long_2023,mo_anomalous_2022,overland_causes_2021,duan_impact_2025,white_unprecedented_2023}. 

We confirmed that TraCE-ST remains stable across a broad range of hyperparameter configurations and can be used with different causal discovery algorithms, although its performance inherits the strengths and weaknesses of the selected algorithm. In our synthetic experiments, Elastic-Net outperformed PCMCI and DYNOTEARS while remaining computationally efficient. Future work should evaluate TraCE-ST with additional time-series causal-discovery algorithms across scientific applications and improve its compatibility with multiple causal discovery engines without sacrificing computational tractability. As with other data-driven causal discovery approaches, the inferred causal pathways are conditional on the variables and spatial domain included in the analysis. Omitted drivers, unresolved processes, confounders, or an insufficiently large spatial domain may therefore produce incomplete or incorrect causal interpretations. Temporally correlated errors, which are common in geoscience data, may also generate spurious causal links. Moreover, the observational reanalyses and numerical simulations provided as inputs to TraCE-ST may contain biases or uncertainties in transport, chemistry, and other physical processes that can propagate into the obtained causal pathways. These limitations motivate careful interpretation and application of TraCE-ST across multiple observational products or Earth system models to assess the robustness of results to observational uncertainty and differences in process representation.

Robust TraCE-ST performance depends on informed variable selection, inclusion of relevant processes at scales resolvable by the input data, and physically defensible choices for the temporal window length, search region size, and causal stencil geometry. These choices constitute part of the model specification, rather than incidental tuning parameters, and their influence should be assessed through sensitivity analyses. The physical principles used here to define admissible hyperparameter ranges could also provide a basis for selecting hyperparameters across different Earth system applications. Several extensions could further improve the framework, including adaptive time windows, traced regions that evolve in shape rather than remaining rectangular, and multi-lag or multiscale causal estimation. These developments could better represent causal structures that are filamentary, deformable, or slowly evolving.

Our results motivate several broader uses. TraCE-ST can serve as a hypothesis-generation tool that narrows the set of candidate precursor regions, variables, and processes before more expensive numerical experimentation, a capability that is particularly useful for rare and undersampled extremes \cite{osman_globally_2021,miloshevich_probabilistic_2023,katzenberger_developing_2026}. When applied to sufficiently large event sets, TraCE-ST could also support comparisons of causal pathways inferred from reanalyses, observations, and numerical, hybrid, or AI-based Earth system models, serving as a process-based evaluation \cite{fan_performance_2020,karmouche_regime-oriented_2023,galytska_evaluating_2023,benard_causality-based_2025,simpson_confronting_2025,henn_aimip_2026}. TraCE-ST could also help address outstanding questions in weather and climate science. For example, it could clarify the relative roles of environmental drivers and inner-core processes in tropical cyclone rapid intensification \cite{smith_toward_2015}. It could also complement recent machine learning-based studies of the sources and limits of Earth system predictability by identifying the relative contributions of individual Earth-system components and mechanisms \cite{meehl_initialized_2021,perez-carrasquilla_earth-system-oriented_2025,vonich_atmospheric_2026}. Similarly, applying TraCE-ST across past, present, and future climate states could also help determine whether extremes arise through different causal pathways, providing insight into the respective influences of externally forced change and internal variability \cite{solomon_distinguishing_2011}. Furthermore, since TraCE-ST traces causal pathways without imposing predefined, field-specific thresholds, it provides a complementary methodology to conventional feature-tracking methods \cite{ullrich_tempestextremes_2021,robledo_climatological_2024,feng_mesoscale_2025}.

Data-driven causal tracking does not replace physical reasoning or numerical experimentation; rather, it complements these approaches by providing a systematic, event-conditioned framework for identifying where, when, and through which variables causality is most plausibly propagated. Therefore, we view TraCE-ST as an innovative and useful addition to the current set of scientific tools available for studying Earth system phenomena and extremes.

\section*{Materials and Methods}\label{sec:methods}

\subsection*{M-CaStLe}

To identify the causal parents of the current child $C^{(j)}(\mathcal{R},t)$ at each TraCE-ST iteration over the region of interest $\mathcal{R}$ during the time window $\mathcal{W}$, TraCE-ST relies on the multivariate Causal Space-Time Stencil Learning framework (M-CaStLe) \cite{nichol_mcastle_2026}. M-CaStLe enables causal discovery in gridded spatiotemporal data under sample-limited conditions, where the number of spatial degrees of freedom often exceeds the number of temporal observations available within an event-focused window. The framework increases the sample size by exploiting spatial regularity and locality in causal relationships. Specifically, it relies on four key assumptions: (i) temporal locality, whereby causal dependencies are restricted to a single lag ($t-1 \rightarrow t$); (ii) spatial locality, whereby causal parents of a grid cell lie within a finite-radius neighborhood; (iii) temporal stationarity, whereby causal relationships are approximately constant within the time window $\mathcal{W}$; and (iv) spatial stationarity, whereby these relationships are shared across locations within the analysis region $\mathcal{R}$. These assumptions should be interpreted as approximations rather than exact requirements, and their validity depends on selecting analysis regions and time windows over which the dominant physical processes evolve smoothly enough for neighboring samples to provide meaningful information about the underlying causal structure. As these assumptions become progressively less valid, uncertainty in the obtained causality generally increases. Consequently, the hyperparameter ranges explored by TraCE-ST must be limited to conditions under which these assumptions are expected to hold approximately.

Under these assumptions, M-CaStLe converts the data within $\mathcal{R}$ and $\mathcal{W}$ into a locally encoded dataset. For each timestep $t \in \mathcal{W}$ and valid spatial location $s \in \mathcal{R}$, it extracts the $t-1$ lagged values of all variables from the $(2r+1)\times(2r+1)$ Moore neighborhood $\mathcal{B}_r(s)$. To ensure that each stencil is fully contained within $\mathcal{R}$, central locations must lie at least $r$ grid cells from the region boundaries. For a square region of size $n \times n$, this restriction yields $(n-2r)^2$ admissible central locations. Pooling the lagged neighborhood values across central locations and time steps produces a design matrix containing $N_{\text{samples}} = |\mathcal{W}| \times (n-2r)^2$ samples and $N_{\text{features}} = p \times (2r+1)^2$ features, where each sample represents a local spatiotemporal context and each feature corresponds to a particular variable and relative position within the stencil $\mathcal{B}_r(s)$. This transformation effectively converts the original spatiotemporal field into a dataset that draws on both spatial replication and temporal evolution, substantially increasing the number of samples available for causal estimation while constraining the dimensionality of the candidate parent set.

Causal parents are then identified by applying a time series causal-discovery or sparse predictive-modeling method to the resulting design matrix. For each central node $X^{(j)}(s,t)$, candidate predictors include all variables $X^{(a)}(s',t-1)$, where $s' \in \mathcal{B}_r(s)$ and $a \in \{1,\dots,p\}$. We explored three causal discovery engines: Elastic-Net regression \cite{zou_regularization_2005}, DYNOTEARS \cite{pamfil_dynotears_2020}, and PCMCI \cite{runge_detecting_2019}. These methods provide complementary approaches based on structural equation modeling, conditional-independence testing, and sparse predictive modeling, respectively. M-CaStLe returns a localized directed graph of the selected or statistically significant $t-1 \rightarrow t$ relationships within the stencil $\mathcal{B}_r(s)$ (e.g., see Fig. \ref{fig:trace_st_schematic}C,D). These directed graphs define the candidate causal parents that TraCE-ST uses to transition between variables and determine the spatial displacement of $\mathcal{R}$ at each iteration. Because M-CaStLe re-estimates these directed graphs at each TraCE-ST iteration using the current analysis region and temporal window, the locality and stationarity assumptions apply only to the current tracing step. Longer-range pathways can therefore emerge through successive local causal parents.

\subsection*{Clustering and selection of causal parents}

After identifying the lagged parents of the current child variable $C^{(j)}(\mathcal{R},t)$, TraCE-ST uses the DBSCAN (Density-Based Spatial Clustering of Applications with Noise) \cite{schubert_dbscan_2017} to group them by direction. This step distinguishes among causal parents arriving from different directions. Clustering is performed relative to the center of the stencil $\mathcal{B}_r(s)$: each parent is represented by the direction of its spatial offset from the center node, independent of the offset magnitude. Thus, parents with different distances but similar directions can be assigned to the same cluster. For each parent $X^{(a)}(s',t-1)$ of the child $C^{(j)}(\mathcal{R},t)$, let $(\Delta x,\Delta y)$ denote the spatial offset of $s'$ from the center of $\mathcal{B}_r(s)$. Each nonzero offset is projected onto the unit circle and represented by the directional vector
\begin{equation}
(u,v)=
\left(
\frac{\Delta x}{\sqrt{\Delta x^2+\Delta y^2}},
\frac{\Delta y}{\sqrt{\Delta x^2+\Delta y^2}}
\right), \qquad (\Delta x,\Delta y)\neq (0,0)
\end{equation}

DBSCAN is then applied to the directional coordinates $(u,v)$. Clustering is performed separately for each parent variable and, when the causal-discovery engine returns signed relationships, for positive and negative inferred relationships between the causal parent and the causal child so that causal relationships of opposite signs are not mixed. For a prescribed neighborhood radius $\epsilon$ and minimum number of samples $N_{\mathrm{min}}$, DBSCAN identifies dense directional groupings without requiring the number of clusters to be specified a priori. Parents classified as noise by DBSCAN are retained as singleton clusters, allowing isolated but potentially important causal parents to remain eligible for selection as the ``winning cluster.'' Center-to-center causal parents, which have zero spatial displacement and therefore no defined direction, are also retained as singleton clusters.

Once the clusters are defined, TraCE-ST selects a ``winning cluster'' that determines both the child variable and the displacement of the region $\mathcal{R}$ at the next iteration. Let cluster $c$ contain $n_c$ members with absolute causal magnitudes $|\beta_i|$. We compute a cluster strength $S_c$ using either the sum or the mean of the absolute causal magnitudes within the cluster:
\begin{equation}
S_c =
\sum_{i \in c} |\beta_i|
\qquad \text{or} \qquad
S_c =
\frac{1}{n_c}\sum_{i \in c} |\beta_i|
\end{equation}

In deterministic mode, TraCE-ST selects the winning cluster with the largest $S_c$. In probabilistic mode, clusters are sampled according to probabilities derived from their strengths. Two probability rules are supported. Under the linear rule,
\begin{equation}
P(c)=\frac{S_c}{\sum_{c'} S_{c'}}
\end{equation}
\noindent whereas under the softmax rule \cite{bishop_pattern_2006},
\begin{equation}
P(c)=\frac{\exp(\beta_{\mathrm{softmax}} S_c)}{\sum_{c'} \exp(\beta_{\mathrm{softmax}} S_{c'})}
\end{equation}
Here, $\beta_{\mathrm{softmax}}$ controls the degree to which selection probabilities concentrate on stronger clusters. Because cluster strength depends on absolute causal magnitudes, negative and positive relationships are weighted symmetrically during cluster selection.

After a winning cluster $c^\star$ has been selected, its associated parent variable becomes the new child variable for the next iteration. The spatial displacement of the region $\mathcal{R}$ is then computed as the weighted mean of the offsets within $c^\star$,
\begin{equation}
(\overline{\Delta x},\overline{\Delta y})
=
\frac{\sum_{i\in c^\star} w_i (\Delta x_i,\Delta y_i)}
{\sum_{i\in c^\star} w_i},
\qquad
w_i = |\beta_i|^{\alpha}
\end{equation}
where $\alpha$ controls the degree to which higher-magnitude parent-child relationships influence the displacement estimate. As tracing proceeds backward in time, the center of $\mathcal{R}$ is shifted by the resulting spatial displacement. At each step, $\mathcal{R}$ is redefined as the set of grid cells within $\pm\frac{1}{2}\,\text{box size}$ of the new, non-rounded center, thereby propagating the trajectory along the selected causal direction.

\subsection*{TraCE-ST hyperparameters}

TraCE-ST can produce consistent trajectories across broad regions of the hyperparameter space, although its outputs remain sensitive to specific parameter choices. As with any data-driven method, successful application therefore requires physically and computationally defensible ranges to explore. TraCE-ST hyperparameters are associated with three components: the tracking procedure, M-CaStLe, and the selected causal discovery algorithm. We describe each of these below, and Table \ref{tab:hyperparameter_ranges} summarizes the hyperparameter ranges explored for each case study in the main text. We selected these hyperparameter ranges to exclude configurations that produced unstable causal discovery, little or no spatial progression of the traced region, or premature trajectory termination. The ranges considered here therefore represent physically and computationally defensible portions of parameter space rather than all possible hyperparameter choices. Within these hyperparameter ranges, we evaluated whether the inferred trajectories, including their variables, spatial pathways, and propagation directions, remained qualitatively consistent across configurations. Our goal was to assess the stability of TraCE-ST trajectories across hyperparameter configurations, rather than to identify a unique optimal configuration. Users should similarly assess and report sensitivity of their results to the hyperparameter ranges considered.

\subsubsection*{Trajectory hyperparameters} 

Before applying TraCE-ST, the user must define the target child by selecting the starting latitude, longitude, time, and variable. The spatial grid spacing and temporal resolution of the input data are particularly important choices. A finer temporal resolution enables faster causal interactions to be resolved; conversely, relevant drivers operating on timescales shorter than the sampling interval may be missed. For example, in the Mount Pinatubo eruption case, TraCE-ST did not resolve the SO$_2$ $\to$ H$_2$SO$_4$ transformation because this reaction occurs on a timescale shorter than the daily temporal resolution of the input data. Spatial grid spacing must also be fine enough to provide adequate sampling within the region of interest. However, if the grid spacing is too fine relative to the process's displacement speed, a much larger stencil radius may be required to capture fast advective motion between consecutive time steps. The Mount Pinatubo eruption case also illustrates this trade-off: at 4$^\circ$ grid spacing, advection of roughly 16$^\circ$ day$^{-1}$ can be captured with a stencil radius of 4 grid cells, whereas at 1$^\circ$ grid spacing, the same displacement would require a stencil radius of 16 grid cells. Such large stencils may become expensive computationally or intractable for the underlying causal discovery algorithm. 

Clustering the causal parents of the current child requires specifying the DBSCAN neighborhood radius $\epsilon$ and the minimum number of samples $N_{\mathrm{min}}$ needed to form a cluster. Increasing $\epsilon$ allows parents with a wider range of directions to be grouped together, but excessively large values may merge distinct causal sources arriving from different directions. Increasing $N_{\mathrm{min}}$ requires greater directional support for a cluster to form, but may cause smaller groups of causal parents to be classified as noise and treated as singleton clusters. In our applications, sensible values were typically $\epsilon = 0.05-0.3$ and $N_{min} = 2-4$. Other parameters control ``winning cluster'' selection and trajectory displacement. The scoring rule determines whether cluster strength is calculated from the mean or sum of the absolute causal magnitudes of its members. The $\alpha$ parameter controls the weights $|\beta_i|^{\alpha}$ assigned to causal parents within the winning cluster when computing its weighted mean offset from the center of the stencil $\mathcal{B}_r(s)$. Larger values of $\alpha$ give greater influence to child-parent relationships with larger absolute magnitudes. In probabilistic mode, the probability rule determines whether cluster-selection probabilities are calculated through linear normalization or a softmax function. In the latter case, a scaling parameter $\beta_{\mathrm{softmax}}$ controls how strongly the selection probabilities concentrate on clusters with greater strength.

\subsubsection*{M-CaStLe hyperparameters} 

The user must also specify the size of the analysis region $\mathcal{R}$, the length of the time window $\mathcal{W}$, and the radius of the stencil $\mathcal{B}_r(s)$. The size of $\mathcal{R}$ and length of $\mathcal{W}$ determine the number of samples available to the causal discovery algorithm after local stencil encoding. A region or time window that is too small may provide insufficient samples for robust causal discovery. Conversely, a region or time window that is too large may combine distinct dynamics, violating the spatial or temporal stationarity assumptions of M-CaStLe. Although the applications presented here use relatively short time windows to study event-specific dynamics, longer windows are equally compatible with the framework provided that the assumption of approximate stationarity remains reasonable over the selected period. The stencil radius $r$ must be large enough to capture the relevant $t-1 \to t$ dependencies affecting the child variable. A radius that is too small may miss advective or spatially displaced processes, whereas a radius that is too large increases the number of candidate parents and can make the causal discovery problem computationally expensive or intractable. More guidance on M-CaStLe hyperparameters can be found in Nichol et al. \cite{nichol_mcastle_2026}.

\subsubsection*{Causal discovery core hyperparameters} 

Any compatible time-series causal-discovery method can be used within M-CaStLe to infer directed child-parent relationships inside the stencil $\mathcal{B}_r(s)$. Each method introduces its own set of hyperparameters that can influence the inferred local graph and, consequently, TraCE-ST trajectories. In the synthetic two-variable case, we evaluated Elastic-Net, PCMCI, and DYNOTEARS. For Elastic-Net, we varied the regularization strength $\lambda_a$ and the $\ell_1$ ratio. For PCMCI, we explored different significance thresholds for both the PC step and the final graph construction $\alpha_{\mathrm{PC}}$. We also evaluated two conditional-independence tests (ParCorr and RobustParCorr), and optional false discovery rate correction using the Benjamini-Hochberg procedure. For DYNOTEARS, we varied the regularization strength $\lambda_a$, the maximum number of iterations (100--3000), and the dependence threshold used to retain inferred child-parent relationships. These ranges were selected to produce causal graphs spanning a range of sparsity levels while maintaining numerical stability and computational tractability.

\begin{table}[t]
\centering
\caption{\textbf{Hyperparameter search ranges explored for each application of TraCE-ST.}
Search spaces sampled in the hyperparameter ensembles described in the main text. Ranges shown represent admissible exploratory domains used for sensitivity characterization rather than exhaustive or optimized search ranges.}
\label{tab:hyperparameter_ranges}

\scriptsize

\begin{tabular}{p{3.1cm}p{2.25cm}p{2.15cm}p{2.05cm}p{2.05cm}p{2.15cm}}
\hline
\textbf{Hyperparameter} &
\shortstack{\textbf{Synthetic}\\ \textbf{two variables}} &
\shortstack{\textbf{Synthetic}\\ \textbf{three variables}} &
\shortstack{\textbf{Debby}\\ \textbf{(2006)}} &
\shortstack{\textbf{Pinatubo}\\ \textbf{(1991)}} &
\shortstack{\textbf{PNW heatwave}\\ \textbf{(2021)}}\\
\hline

\multicolumn{6}{l}{\textit{Trajectory and M-CaStLe hyperparameters}}\\

Time window $\mathcal{W}$ &
$2$--$4$ samples &
$5$--$7$ samples &
$5$--$7$ h &
$3$--$5$ d &
$4$--$6$ d \\

Region size $\mathcal{R}$ (box size) &
$15$--$30$ grid cells&
$30$--$40$ grid cells&
$5$--$7^\circ$ &
$65$--$75^\circ$ &
$30$--$40^\circ$ \\

Stencil radius $r$ &
$2$--$3$ &
$4$--$6$ &
$4$--$6$ &
$5$--$7$ &
$4$--$7$ \\

DBSCAN neighborhood $\epsilon$ &
$0.05$--$0.25$ &
$0.05$--$0.25$ &
$0.05$--$0.25$ &
$0.05$--$0.25$ &
$0.05$--$0.25$ \\

DBSCAN $N_{\min}$ &
$2$ &
$2$ &
$2$ &
$2$ &
$2$ \\

Scoring rule &
mean, sum &
mean, sum &
mean, sum &
mean, sum &
mean, sum \\

$\alpha$ &
$0$--$64$ &
$0$--$64$ &
$0$--$64$ &
$0$--$64$ &
$0$--$64$ \\

\multicolumn{6}{l}{\textit{Probabilistic trajectory hyperparameters}}\\

Probability rule &
linear, softmax &
linear, softmax &
linear, softmax &
linear, softmax &
linear, softmax \\

Softmax scaling $\beta_{softmax}$ &
$0$--$64$ &
$0$--$64$ &
$0$--$64$ &
$0.5$--$64$ &
$0$--$64$ \\

\multicolumn{6}{l}{\textit{Causal discovery core hyperparameters}}\\

Elastic-Net $\lambda_a$ &
$10^{-3}$--$10^{-0.5}$ &
$10^{-2.5}$--$10^{-1}$ &
$10^{-2.5}$--$10^{-1}$ &
$10^{-0.3}$--$1$ &
$10^{-2.5}$--$1$ \\

Elastic-Net $\ell_1$ ratio &
$10^{-4}$--$1$ &
$10^{-1}$--$1$ &
$10^{-1}$--$1$ &
$10^{-0.2}$--$1$ &
$10^{-1}$--$1$ \\

PC significance threshold $\alpha_{PC}$ (PCMCI) &
$\{10^{-4}-10^{-1}\}$ &
--- &
--- &
--- &
--- \\

Graph threshold (PCMCI) &
$\{10^{-4}-10^{-1}\}$ &
--- &
--- &
--- &
--- \\

Conditional independence test (PCMCI) &
ParCorr, RobustParCorr &
--- &
--- &
--- &
--- \\

FDR correction (PCMCI) &
None, BH &
--- &
--- &
--- &
--- \\

DYNOTEARS $\lambda_a$ &
$10^{-3}$--$10^{-0.5}$ &
--- &
--- &
--- &
--- \\

Maximum iterations (DYNOTEARS) &
$10^{2}$--$3\times10^{3}$ &
--- &
--- &
--- &
--- \\

Dependence threshold (DYNOTEARS) &
$10^{-6}$--$10^{-2}$ &
--- &
--- &
--- &
--- \\
\hline
\end{tabular}
\end{table}


\clearpage 




\section*{Acknowledgments}
JSPC thanks Jon Poterjoy, Jim Carton, and Xin-Zhong Liang, who, as part of his PhD committee, provided helpful comments related to this manuscript. Additionally, we acknowledge the use of large language models (ChatGPT 5.3 and Claude Sonnet 4.6) to refine the code and edit parts of the manuscript.

\paragraph*{Funding}

JSPC and MJM were supported by a University of Maryland Grand Challenges Seed Grant. JSPC also acknowledges support from the Dr. Ann G. Wylie Dissertation Fellowship and the Dr. Eugene Rasmusson Graduate Student Fellowship from The Graduate School at the University of Maryland. KD was supported by the Regional and Global Model Analysis (RGMA) component of the Earth and Environmental System Modeling Program of the U.S. Department of Energy’s Office of Biological \& Environmental Research (BER) under Lawrence Livermore National Lab subaward DE-AC52-07NA27344, Lawrence Berkeley National Lab subaward DE-AC02-05CH11231, and Pacific Northwest National Lab subaward DE-AC05-76RL01830, as well as by the National Science Foundation (NSF) National Center for Atmospheric Research (NCAR), which is a major facility sponsored by NSF under Cooperative Agreement No. 1852977. MNE was supported by NSF grant 2303530. VR was supported by a NASA FINESST grant 80NSSC25K0649. JJN and DB were supported by the Laboratory Directed Research and Development program at Sandia National Laboratories, a multimission laboratory managed and operated by National Technology and Engineering Solutions of Sandia, LLC, a wholly owned subsidiary of Honeywell International, Inc., for the U.S. Department of Energy’s National Nuclear Security Administration under contract DE-NA-0003525. This written work is authored by employees of NTESS. The employees, not NTESS, own the right, title, and interest in and to the written work and are responsible for its contents. Any subjective views or opinions that might be expressed in the written work do not necessarily represent the views of the U.S. Government. The publisher acknowledges that the U.S. Government retains a non-exclusive, paid-up, irrevocable, world-wide license to publish or reproduce the published form of this written work or allow others to do so, for U.S. Government purposes. The DOE will provide public access to results of federally sponsored research in accordance with the DOE Public Access Plan. This paper describes objective technical results and analysis. Any subjective views or opinions that might be expressed in the paper do not necessarily represent the views of the U.S. Department of Energy or the United States Government. Computing and data storage resources were provided by the Computational and Information Systems Laboratory at the NSF NCAR.

\paragraph*{Author contributions}

JSPC, JJN, VR, DB, and MJM conceptualized the study. JSPC developed the methodology and software, designed and performed the experiments, conducted the formal analysis, developed the visualizations, and wrote the original draft of the manuscript. JJN, VR, DB, KD, MNE, and MJM contributed to the development of the methodology and the interpretation of results. JSPC, JJN, and VR curated the data. MJM supervised the project. All authors contributed to the review and editing of the manuscript and approved the final version.  

\paragraph*{Competing interests}

The authors declare that they have no competing interests.

\paragraph*{Data and materials availability}

All data and code necessary to reproduce the results in this study are publicly available. The TraCE-ST framework, including the tracking algorithm implementation, analysis scripts, and tutorials, is available on GitHub (\url{https://github.com/jhayron-perez/trace-st}) \cite{trace_st_code}. Synthetic datasets for the controlled experiments can be generated with the scripts provided in the repository \cite{trace_st_code}. For the real-world case studies, all input datasets are publicly accessible. Precipitation data were obtained from the GPM IMERG Final product (V07) \cite{huffman_integrated_2020}. Infrared brightness temperature data were obtained from the NCEP/CPC merged IR dataset (MERGIR V1) \cite{janowiak2017ncep}. Atmospheric reanalysis data (ERA5) were obtained from the NCAR Geoscience Data Exchange (GDEX) \cite{era5_dataset}. Simulation data for the Mount Pinatubo experiment correspond to the E3SMv2-SPA configuration described in Brown et al. \cite{brown_validating_2024}. All processed data, trajectory ensembles, and intermediate outputs used to generate the figures in this study are archived at Zenodo \cite{trace_st_data}.


\subsection*{Supplementary materials}
Figures S1--S9


\newpage


\renewcommand{\thefigure}{S\arabic{figure}}
\renewcommand{\thetable}{S\arabic{table}}
\renewcommand{\theequation}{S\arabic{equation}}
\renewcommand{\thepage}{S\arabic{page}}
\setcounter{figure}{0}
\setcounter{table}{0}
\setcounter{equation}{0}
\setcounter{page}{1} 


\begin{center}
\section*{Supplementary Materials for\\ \scititle}

Jhayron S. Pérez-Carrasquilla$^{1\ast}$,
J. Jake Nichol$^{2}$,
Vanessa Robledo$^{3},$\\
Diana Bull$^{2}$,
Katherine Dagon$^{4}$,
Michael N. Evans$^{5,6}$,
and Maria J. Molina$^{1}$\\
\small$^\ast$Corresponding author. Email: jhayron@umd.edu\\
\end{center}

\subsubsection*{This PDF file includes:}
Figures S1--S9
\newpage


\begin{figure} 
	\centering
	\includegraphics[width=0.5\textwidth]{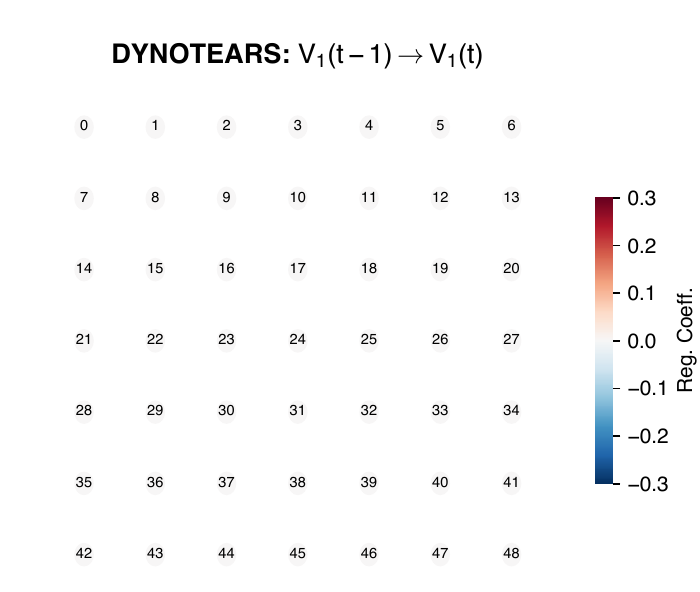} 

	\caption{\textbf{DYNOTEARS results for the candidate $V_1(t-1)\to V_1(t)$ relationship.}
		Same as Figure \ref{fig:synth1}D, with $V_1$ as the candidate causal parent. DYNOTEARS identifies no $V_1(t-1)\to V_1(t)$ relationships in this example.}
	\label{fig:sup_dyno} 
\end{figure} 

\begin{figure} 
	\centering
	\includegraphics[width=0.8\textwidth]{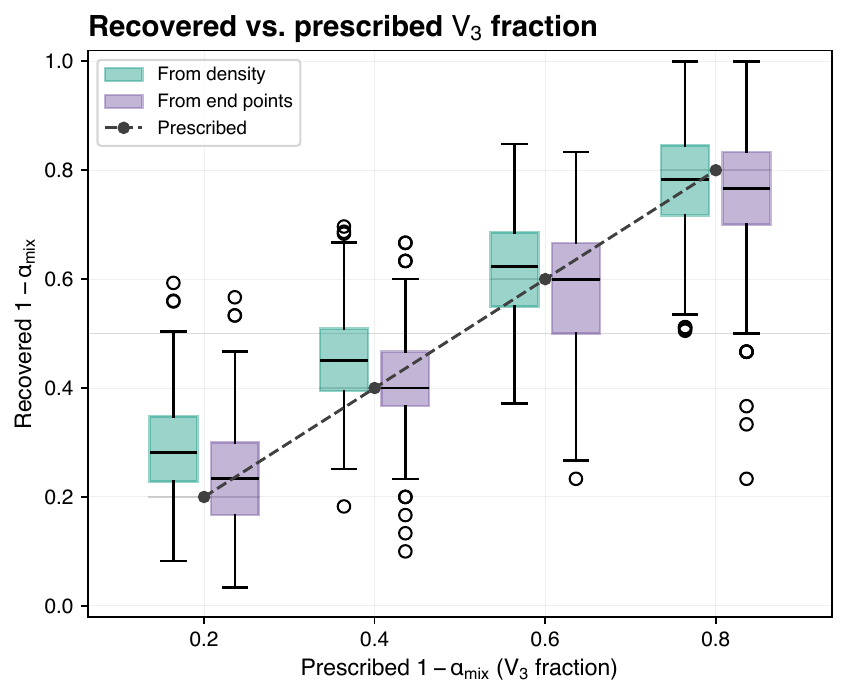} 

	\caption{\textbf{Prescribed vs. recovered $1-\alpha_{mix}$ for synthetic experiment.}
		Same as Figure \ref{fig:synth2}B but using the $V_3$ fraction.}
	\label{fig:sup_alpha} 
\end{figure}

\begin{figure} 
	\centering
	\includegraphics[width=0.8\textwidth]{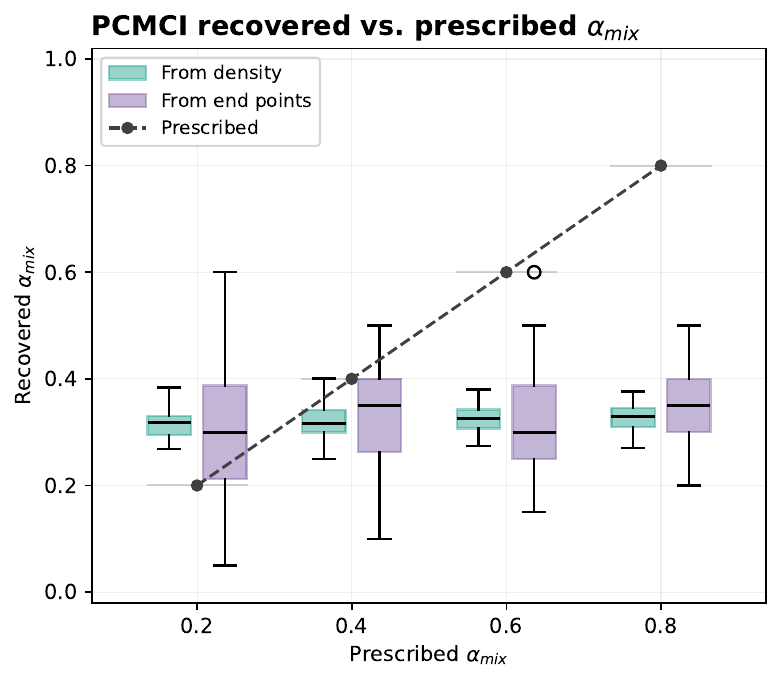} 

	\caption{\textbf{Prescribed vs. recovered $\alpha_{mix}$ for synthetic experiment using PCMCI.}
		Same as Figure \ref{fig:synth2}B but using PCMCI as the underlying causal discovery algorithm.}
	\label{fig:sup_alpha_pcmci} 
\end{figure}

\begin{figure} 
	\centering
	\includegraphics[width=1\textwidth]{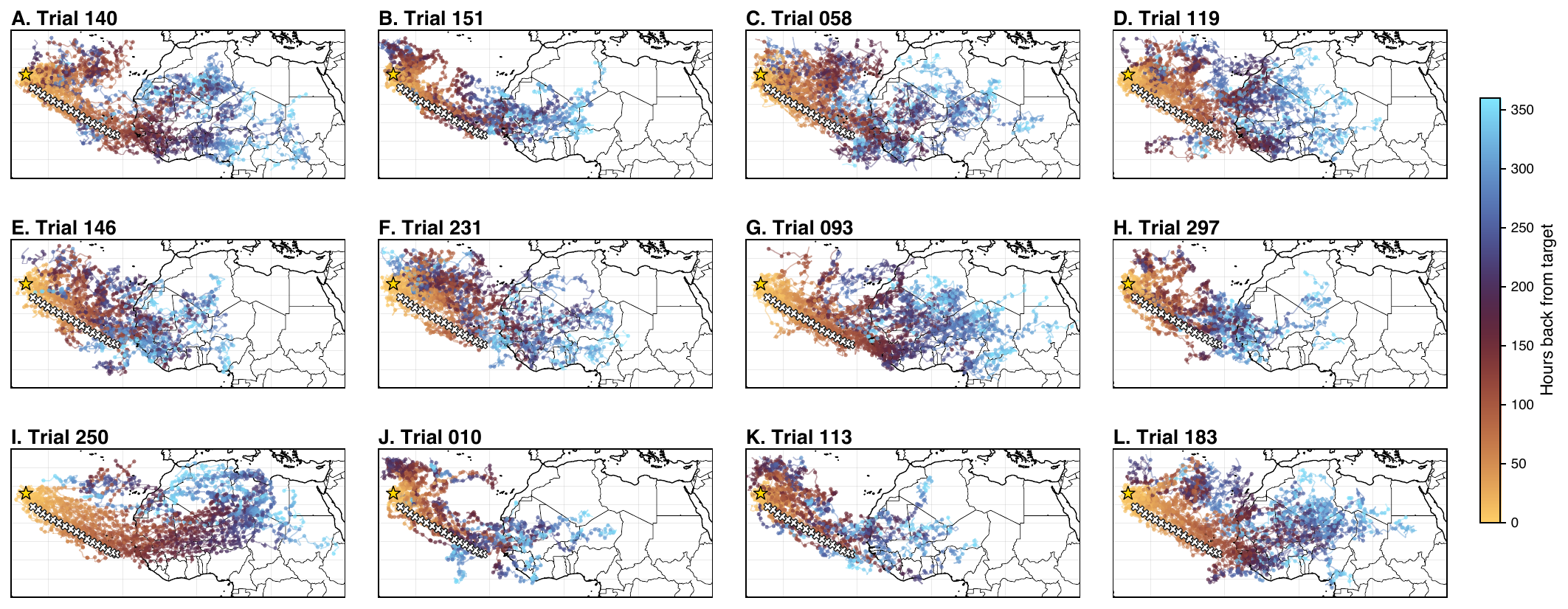} 

	\caption{\textbf{Random examples of TraCE-ST ensembles for Debby (2006).}
		Same as Figure \ref{fig:debby}A for 12 randomly chosen ensembles from the 68 analyzed.}
	\label{fig:sup_ens_debby} 
\end{figure}

\begin{figure} 
	\centering
	\includegraphics[width=0.8\textwidth]{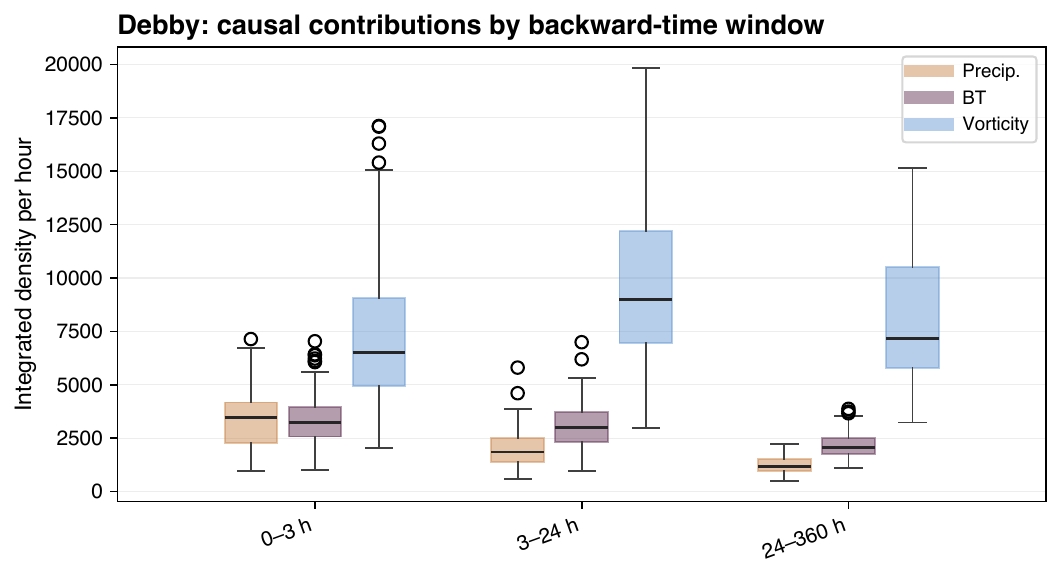} 

	\caption{\textbf{Evolution of causal contributions for Debby (2006).}
		Boxplots of estimated causal contributions, defined as integrated trajectory density, during different periods prior to the target child.}
	\label{fig:sup_dens_debby} 
\end{figure}

\begin{figure} 
	\centering
	\includegraphics[width=1\textwidth]{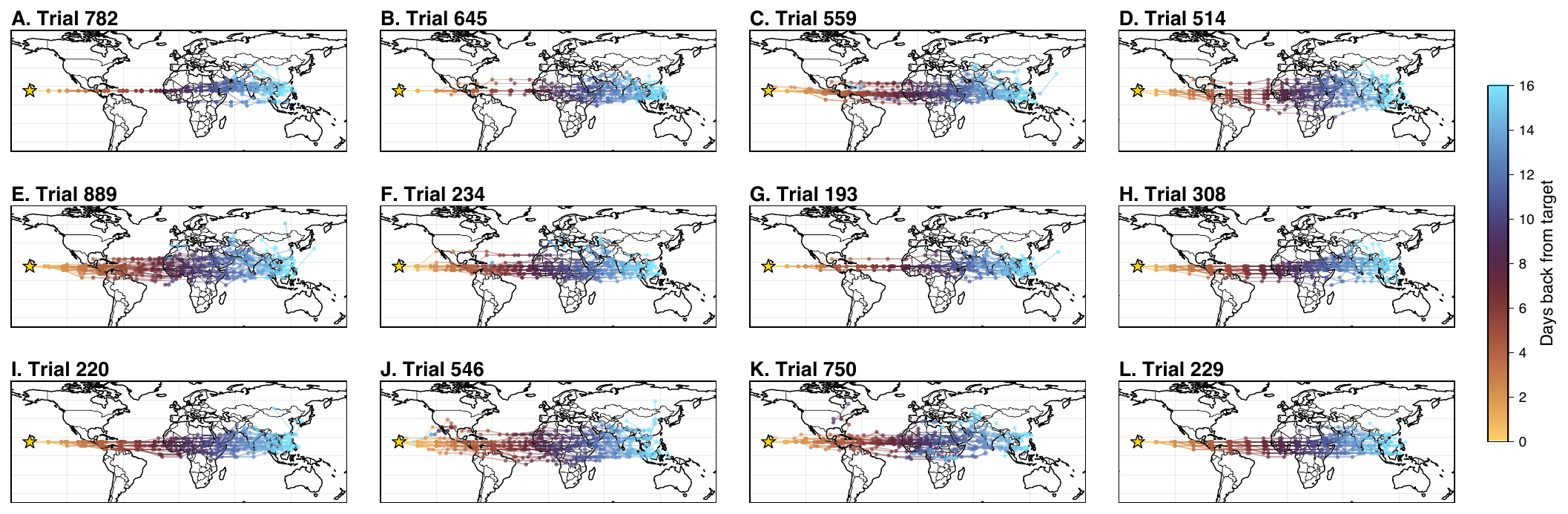} 

	\caption{\textbf{Random examples of TraCE-ST ensembles for Mount Pinatubo Eruption (1991).}
		Same as Figure \ref{fig:pinatubo}A for 12 randomly chosen ensembles from 1000 analyzed.}
	\label{fig:sup_ens_pinatubo} 
\end{figure}

\begin{figure} 
	\centering
	\includegraphics[width=0.8\textwidth]{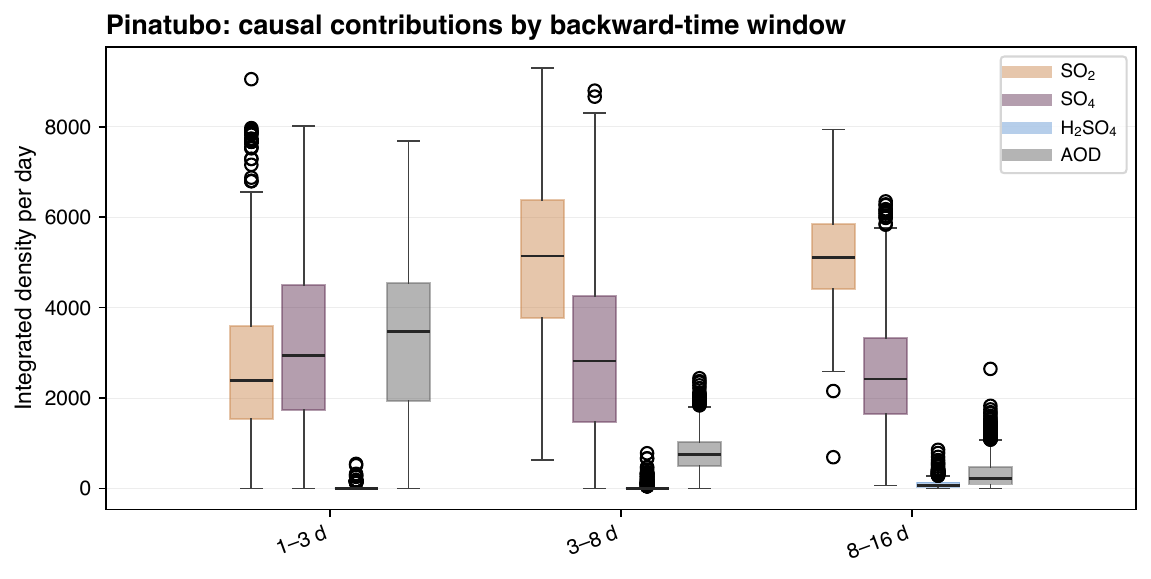} 

	\caption{\textbf{Evolution of causal contributions for Mount Pinatubo Eruption (1991).}
		Boxplots of estimated causal contributions, defined as integrated trajectory density, during different periods prior to the target child.}
	\label{fig:sup_dens_pinatubo} 
\end{figure}

\begin{figure} 
	\centering
	\includegraphics[width=1\textwidth]{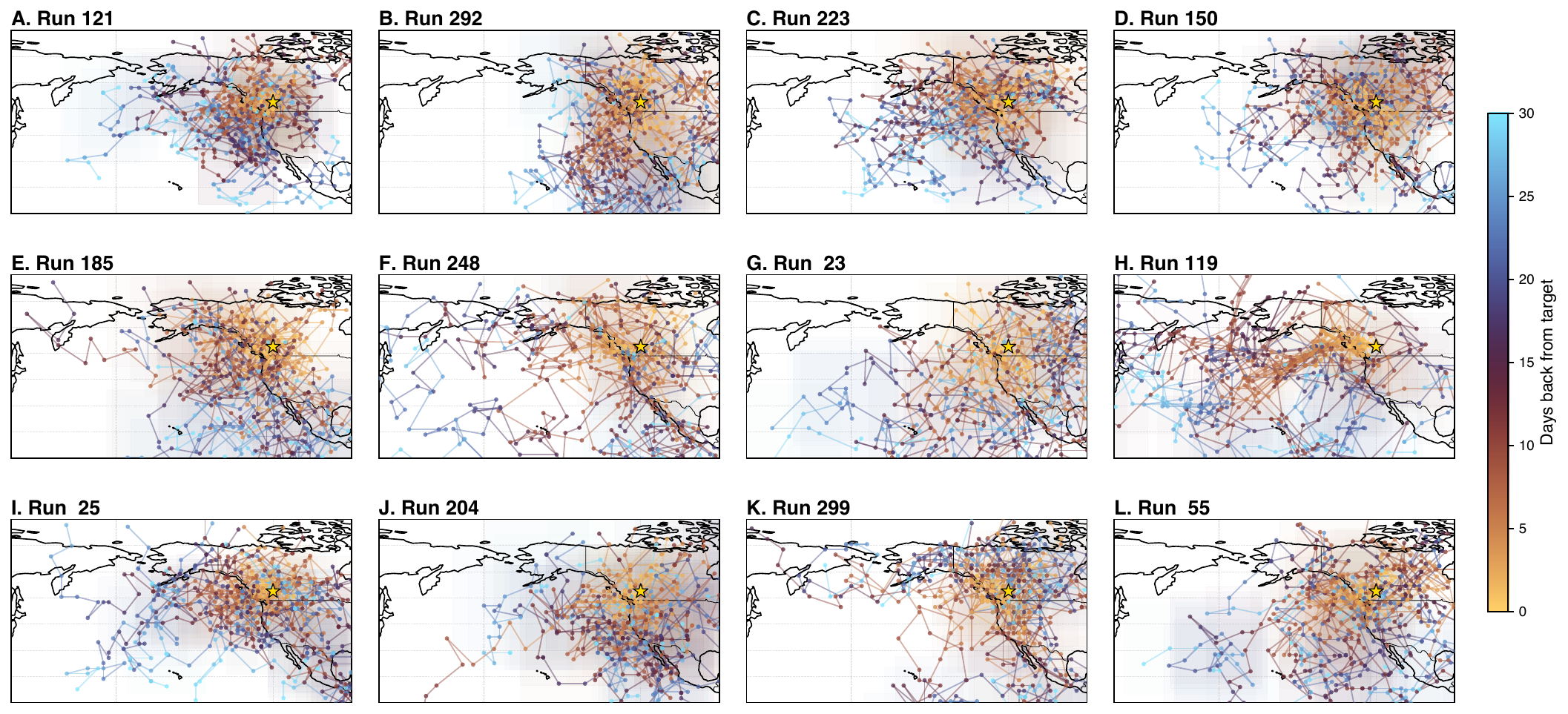} 

	\caption{\textbf{Random examples of TraCE-ST ensembles for the PNW heatwave (2021).}
		Same as Figure \ref{fig:pnw1}A for 12 randomly chosen ensembles from the 242 analyzed.}
	\label{fig:sup_ens_pnw21} 
\end{figure}

\begin{figure} 
	\centering
	\includegraphics[width=1\textwidth]{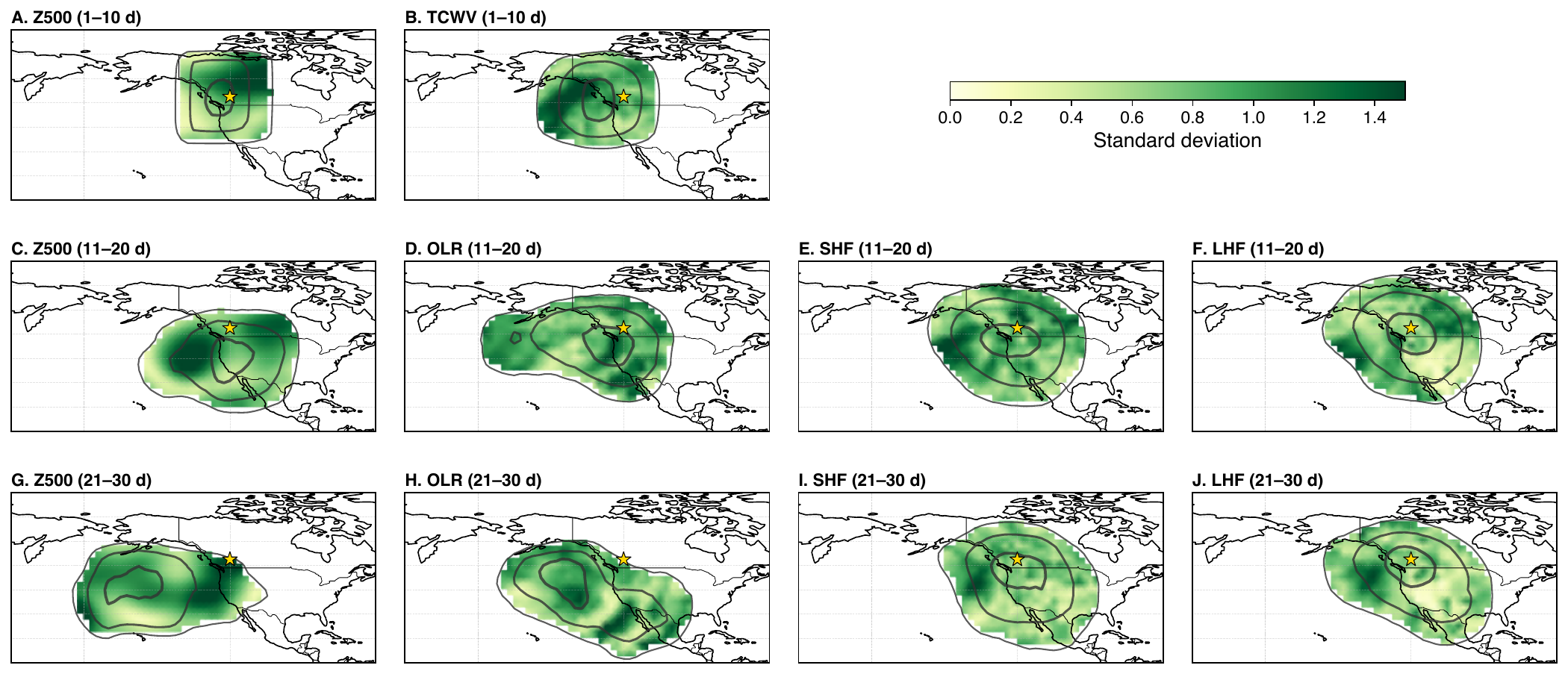} 

	\caption{\textbf{Relevant anomaly patterns identified by TraCE-ST for the 2021 PNW heatwave.}
		Same as Figure \ref{fig:pnw2} but showing the standard deviation instead of the mean.}
	\label{fig:sup_pnw21_std} 
\end{figure}


\end{document}